\documentclass[aps,pra,10pt,twocolumn,superscriptaddress,floatfix]{revtex4-2}
\usepackage{graphicx}
\usepackage{amsmath}
\allowdisplaybreaks[1]
\usepackage{booktabs}
\usepackage{hyperref}
\usepackage{amssymb}
\usepackage[T1]{fontenc}

\graphicspath{{./}}
\newcommand{\ket}[1]{| #1 \rangle}
\newcommand{\bra}[1]{\langle #1 |}
\newcommand{\paren}[1]{\left( #1 \right)}

\begin{document}

\title{Hybrid quantum repeater chains with atom-based quantum processing units and quantum memory multiplexers}

\author{Shin Sun}
\affiliation{Experimental Quantum Information Physics Unit, Okinawa Institute of Science and Technology, Onna, Okinawa 904-0495, Japan}
\affiliation{Networked Quantum Devices Unit, Okinawa Institute of Science and Technology, Onna, Okinawa 904-0495, Japan}
\affiliation{LQUOM, Inc., 79-5 Tokiwadai, Hodogaya-ku, Yokohama, Kanagawa 240-8501, Japan}

\author{Daniel Bhatti}
\affiliation{Networked Quantum Devices Unit, Okinawa Institute of Science and Technology, Onna, Okinawa 904-0495, Japan}

\author{Shaobo Gao}
\affiliation{Experimental Quantum Information Physics Unit, Okinawa Institute of Science and Technology, Onna, Okinawa 904-0495, Japan}

\author{David Elkouss}
\affiliation{Networked Quantum Devices Unit, Okinawa Institute of Science and Technology, Onna, Okinawa 904-0495, Japan}

\author{Hiroki Takahashi}
\email[Contact author: ]{hiroki.takahashi@oist.jp}
\affiliation{Experimental Quantum Information Physics Unit, Okinawa Institute of Science and Technology, Onna, Okinawa 904-0495, Japan}

\date[]{Received 25 December 2025; revised 28 June 2026; accepted 5 August 2026; published 20 October 2026}

\begin{abstract}
Quantum repeaters enable the generation of reliable entanglement across long distances despite the underlying channel noise. Nevertheless, realizing quantum repeaters poses a difficult engineering challenge due to various device constraints and design tradeoffs. Herein we propose and analyze an efficient hybrid quantum repeater design that integrates atom-based quantum processing units, spontaneous parametric down-conversion photon sources, and atomic frequency comb quantum memories. Our design leverages the strong spectrotemporal multiplexing capability of the quantum memory to enable high-rate elementary-link entanglement generation between repeater nodes. Transferring the photonic entanglement into matter-qubit entanglement, together with deterministic quantum operations, further enables reliable long-distance entanglement distribution. We analyze photon-loss channels in the hybrid architecture and propose suitable error-suppression strategies that are natively incorporated into our repeater protocol. Using numerical simulations, we demonstrate the advantages of our hybrid design for end-to-end secret key rates in a linear repeater-chain model. With continued advances in relevant hardware technologies, we envision that the proposed hybrid design is well suited for large-scale quantum networks.
\end{abstract}

\maketitle

\section{Introduction}\label{sec:introduction}

Distributing high-fidelity entanglement among distant users is central to a wide range of quantum information processing tasks~\cite{kimble2008quantum,wehner2018quantum}, including blind quantum computing~\cite{broadbent2009universal,barz2012demonstration}, entanglement-based quantum key distribution~\cite{ekert1991quantum,ribordy2000long,yin2017satellite}, and modular architectures for scalable quantum computing~\cite{cirac1999distributed,van2010distributed,monroe2014large}. Experimentally, several platforms for prototypical remote-entanglement distribution have been demonstrated, including ion traps~\cite{stephenson2020high,krutyanskiy2023entanglement}, neutral atoms~\cite{covey2023quantum}, nitrogen-vacancy (NV) centers~\cite{bernien2013heralded,kalb2017entanglement}, and photonic systems~\cite{yu2020entanglement,liu2021heralded}. In practice, due to the inevitable loss of quantum information during long-distance transmission, quantum repeaters~\cite{briegel1998quantum,munro2015inside,azuma2023quantum} are essential for establishing reliable remote entanglement in a scalable way. Nevertheless, the limited repetition rates and the inherently probabilistic nature of photonic entanglement swapping employed in first-generation repeater designs~\cite{munro2015inside} severely constrain both the achievable distance and the scalability of near-term quantum repeater networks.

In order to improve the performance of the first-generation quantum repeater architecture, several previous proposals considered the joint usage of devices with different characteristics to build a hybrid-device quantum repeater~\cite{askarani2021entanglement,gu2024hybrid,cussenot2025uniting}. In~\cite{askarani2021entanglement}, a multiplatform repeater chain employing NV-center-based quantum processing units (QPUs), spontaneous parametric down-conversion (SPDC) sources, atom-based quantum buffers, and atomic frequency comb (AFC) quantum memories (QMs)~\cite{afzelius2009multimode,sinclair2014spectral} was proposed, with distinct multiplexing strategies used across different segments. In their protocol, photonic entanglement is generated by SPDC sources, with one photon stored in the AFC memory and the other transmitted to a remote-entanglement swapper; QM-QM entanglement is then established via a double-click protocol~\cite{marcikic2002time}, where two consecutive clicks at the photon detectors herald the generation of remote entanglement. Loading into matter qubits occurs in two stages: Photonic entanglement is first transferred to a quantum buffer and then swapped into the electron spin of an NV center; subsequently, the electron spin state is transferred to a nuclear spin to free the electronic qubit for further loading. Deterministic routing and merging of elementary links (ELs) are finally performed using the NV-center-based QPUs.

In~\cite{gu2024hybrid}, photon sources based on rubidium atoms were employed to generate photonic entanglement. As in the previous work, QM-QM entanglement is created via a double-click protocol, and the resulting photonic state is subsequently loaded into a cavity-based atomic system for deterministic processing. However, since atom-based photon sources operate at low repetition rates and lack spectral multiplexing, the multiplexing capability is restricted to the temporal domain.

More recently, Ref.~\cite{cussenot2025uniting} proposed another hybrid quantum network architecture. Their quantum network employs ion-trap QPUs as end nodes and an SPDC-based backbone consisting of multiple SPDC sources and QMs that generate dual-rail-encoded photonic entanglement. A carefully shaped AFC readout pulse is used to ensure indistinguishability between SPDC photons and QPU-emitted photons, enabling high-fidelity entanglement transfer from the photonic backbone to the ion-trap end nodes.

In this work, although we follow a similar philosophy of combining photonic and matter-based devices, we introduce a hybrid architecture that offers a simpler experimental implementation and substantially enhanced multiplexing capability. The hybrid repeater design we propose uses SPDC sources, atomic frequency comb quantum memories (AFCQMs), and cavity-coupled atom-based QPUs using ion traps~\cite{takahashi2020strong} or neutral atoms~\cite{li2024high,sinclair2025fault}. On the photonic side, SPDC sources enable the rapid generation of entangled photons with extensive multiplexing capability that are compatible with near-term QMs~\cite{lago2021telecom}. On the atom-based side, cavity-coupled QPUs provide highly efficient light-matter interfaces and support deterministic quantum information processing. The main distinguishing features of our design are as follows.

First, we employ the single-click protocol~\cite{hermans2023entangling} for the remote-entanglement swapping, which requires only one detector click at the entanglement swapper and each of the local repeater units. The single-click protocol has better scaling than the double-click protocol for the success probability due to the fact that it does not require the simultaneous arrival of photons from two arms connecting to the entanglement swapper. This is in contrast with the protocol used in Ref.~\cite{cussenot2025uniting}, where double clicks at each end node are required to load the entanglement from the repeater chains to the QPUs.

Second, our scheme can use both spectral and temporal multiplexing for elementary-link generation, achieving a substantially higher multiplexing capacity than schemes that rely solely on temporal multiplexing. By introducing a QM as a multiplexing module, our hybrid repeater achieves higher entanglement distribution rates over long distances than architectures that use the QPUs themselves as photon sources. The issue of allocating multiplexing resources is investigated with numerical calculations.

Third, as already noted in Ref.~\cite{cussenot2025uniting}, the bandwidth mismatch between the atomic emission and the SPDC output poses problems for the photon indistinguishability required for entanglement swapping. To solve this problem, we propose using cavity-enhanced SPDC sources~\cite{lago2021telecom,bao2008generation} and cavity-coupled atom-based QPUs~\cite{takahashi2020strong,sinclair2025fault}. By carefully engineering both cavities, the bandwidths (and temporal profiles) of the photons from both emitters can be on the same order. Furthermore, pulse shaping techniques either on the atom-based QPU~\cite{keller2004continuous} or on the SPDC source~\cite{ansari2018heralded} can be utilized to refine the temporal overlaps. This choice of cavity-enhanced SPDC directly informs a proper repetition rate for the SPDC photon generation, as we discuss in the numerical results of the paper. The proposed design simplifies the interference conditions between the SPDC photons and atom-emitted photons, reducing the need for specialized AFC readout pulses such as those proposed in Ref.~\cite{cussenot2025uniting}.

Fourth, in our design, photonic entanglement is first generated remotely in a heralded manner and only then transferred to the QPUs. Once loading is complete, the resulting QPU-QPU entanglement is immediately available for deterministic operations. This separation of stages simplifies network scheduling and avoids the multistep loading procedures used in~\cite{askarani2021entanglement}.

We quantitatively demonstrate the advantages of our hybrid repeater through numerical simulations using estimated hardware parameters.

The paper is structured as follows. In Sec.~\ref{sec:background} we review key concepts of the quantum repeater design, including several first-generation quantum repeater proposals. In Sec.~\ref{sec:architecture} we describe the architecture of the proposed hybrid repeater and the associated entanglement-distribution protocol. In Sec.~\ref{sec:error_mitigation} we analyze the impact of photon loss and discuss error-mitigation strategies, including hardware choices and entanglement distillation. In Sec.~\ref{sec:results} we present the main numerical results, derive analytical expressions for end-to-end rates, and optimize the multiplexing strategy. We then benchmark our design against a linear repeater chain based solely on atom-based QPUs. In Sec.~\ref{sec:discussion} we outline future research directions and assess the experimental challenges involved in realizing the proposed architecture.

\section{Background}\label{sec:background}

In this section, we briefly review the background of quantum repeater design and introduce the tradeoff of using different schemes for remote-entanglement generation. The advantages of each design are combined in the hybrid repeater we propose.

\subsection{Direct entanglement swapping between atom-based quantum processors}\label{subsec:ion_trap_based}

We first review the atom-based quantum repeater design, comprising atom-based QPUs interconnected by optical fibers and entanglement swappers. This approach can be viewed as a single-emitter variant of the well-known Duan-Lukin-Cirac-Zoller protocol~\cite{duan2001long}. To ensure ideal light-matter interaction and photon collection efficiency, we assume the atomic emitter is trapped within a high-quality optical cavity.

\begin{figure}[tbp]
\centering
\includegraphics[width=\linewidth]{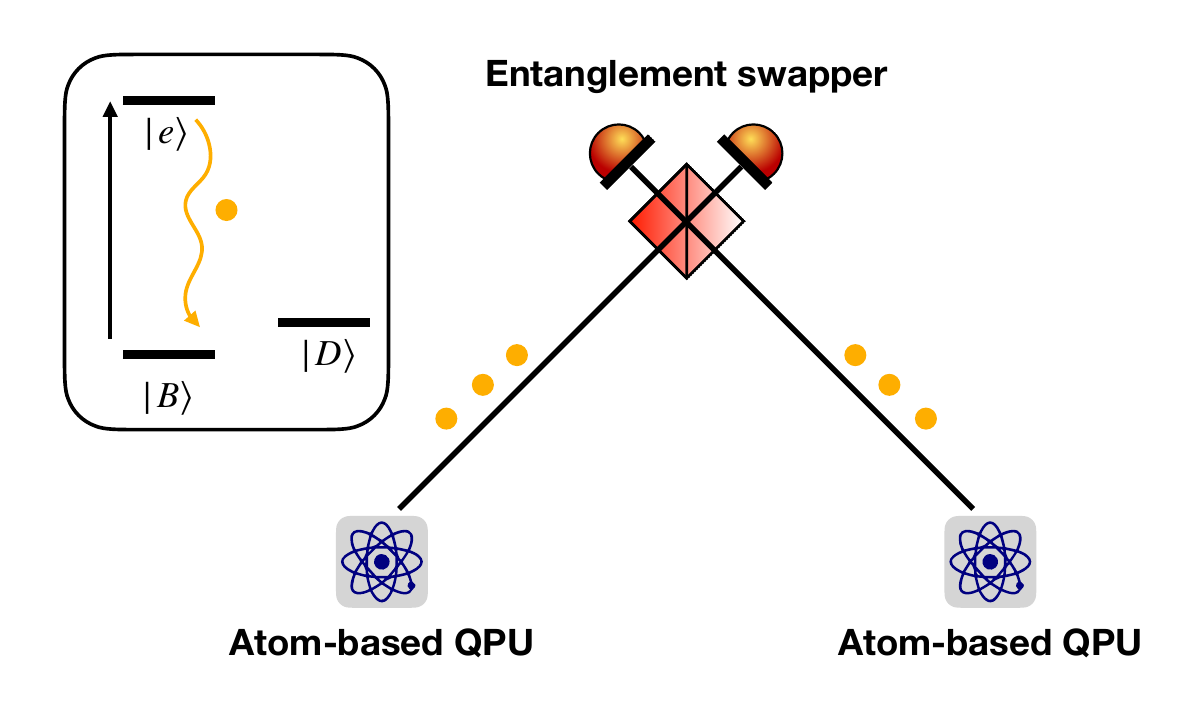}
\caption{Elementary link for atom-based quantum repeaters. The link comprises two atom-based QPUs and an entanglement swapper. Each QPU locally prepares atom-photon entanglement and transmits the photon to a remote swapper. The inset shows an energy diagram of the QPU atom. The cycling transition $\ket{B}\leftrightarrow\ket{e}$ couples to the cavity. A single-click event at the swapper heralds entanglement between remote QPUs.}
\label{fig:atom_based}
\end{figure}

As shown in Fig.~\ref{fig:atom_based}, an elementary link (QPU-QPU entanglement) is generated by preparing the atomic emitter in a superposition of the bright ($\ket{B}$) and dark ($\ket{D}$) ground states. The atom is then excited via the $\ket{B}\rightarrow\ket{e}$ transition (Fig.~\ref{fig:atom_based}, inset), and subsequent state-dependent photon emission yields the atom-photon entangled state~\cite{beukers2024remote}
\begin{align}
\ket{\psi_\text{atom-photon}} = \sqrt{1-q}\ket{D0} + \sqrt{q}\ket{B1}, \label{eqn:ion_photon}
\end{align}
where $0$ and $1$ denote the Fock states of the emitted photon mode. The brightness parameter $q$ is tuned via the initial superposition weight and must be kept small to maintain a high-fidelity link, reflecting the inherent rate-fidelity tradeoff~\cite{hermans2023entangling}.

Emitted photons undergo quantum frequency conversion to the telecommunication band for long-distance transmission before interfering at the entanglement swapper. We employ the single-click entanglement swapping protocol~\cite{cabrillo1999creation} due to its superior rate-distance scaling~\cite{hermans2023entangling}. A successful herald occurs when only one detector behind the central beam splitter clicks (Fig.~\ref{fig:atom_based}), ideally projecting the remote atomic qubits into the Bell state
\begin{align}
|\Psi^+\rangle = \sqrt{\frac{1}{2}}\left(|BD\rangle + |DB\rangle\right). \label{eqn:bell_state}
\end{align}

Due to the probabilistic nature of entanglement swapping and finite signal travel times, QPUs can utilize multiple emitters for temporal multiplexing~\cite{xiong2016active}. Sequentially exciting emitters before receiving the herald reduces waiting times and improves generation rates. Consequently, the repetition rate is upper bounded by the inverse of the atom-photon entanglement generation time, typically in the submegahertz regime~\cite{schupp2021interface}. For extended repeater chains, an optical switch routes photons to entangle a QPU with multiple neighbors. Adjacent elementary links are then merged via deterministic Bell measurements within a QPU, establishing long-range entanglement while circumventing exponential fiber losses.

However, scaling this design remains challenging due to the inherently low repetition rate of atom-photon entanglement generation and the technical difficulty of integrating many emitters with efficient optical interfaces within a single QPU to enhance multiplexing.

\subsection{Quantum repeaters based on SPDC sources and quantum memories}\label{subsec:SPDC}

\begin{figure}[tbp]
\centering
\includegraphics[width=0.8\linewidth]{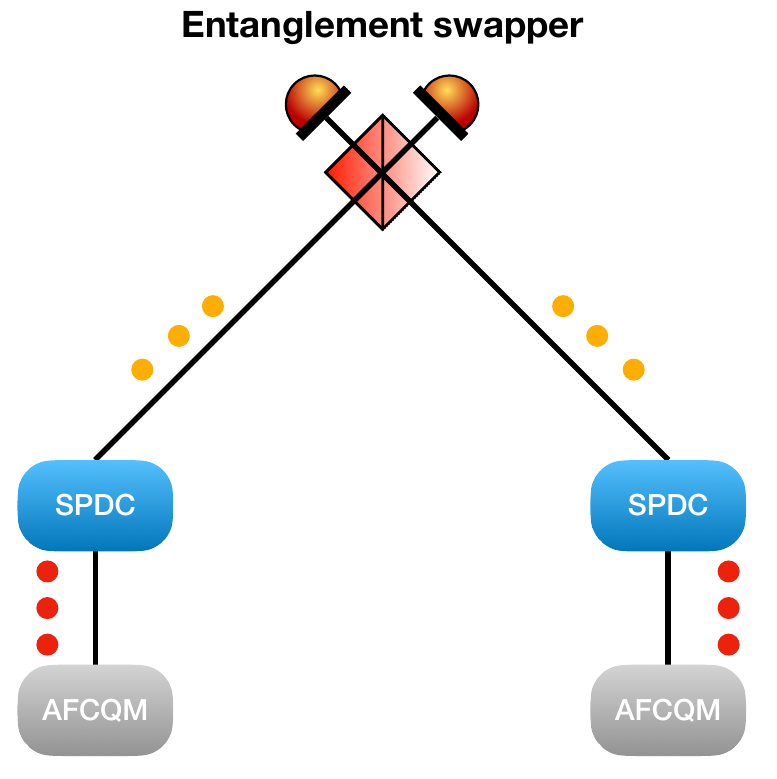}
\caption{Elementary link for SPDC-based quantum repeaters. The SPDC sources are first pumped to generate entangled photon states in the Fock basis. For each entangled state, one mode is sent to a remote-entanglement swapper and another mode is stored in a QM (we assume an AFC-type QM). The center entanglement swapper heralds the generation of entanglement between remote QMs.}
\label{fig:spdc_only}
\end{figure}

In addition to the quantum repeater based on atomic emitters, quantum repeaters based on photonic devices such as SPDC sources and QMs have been proposed (Fig.~\ref{fig:spdc_only})~\cite{lago2021telecom,krovi2016practical}. Quantum repeaters based on SPDC sources typically offer a higher repetition rate and can operate in both temporal and frequency multiplexing schemes. Similar to the atom-based repeaters, we discuss the single-click version of the SPDC-based repeaters.

\begin{figure}[tbp]
\centering
\includegraphics[width=\linewidth]{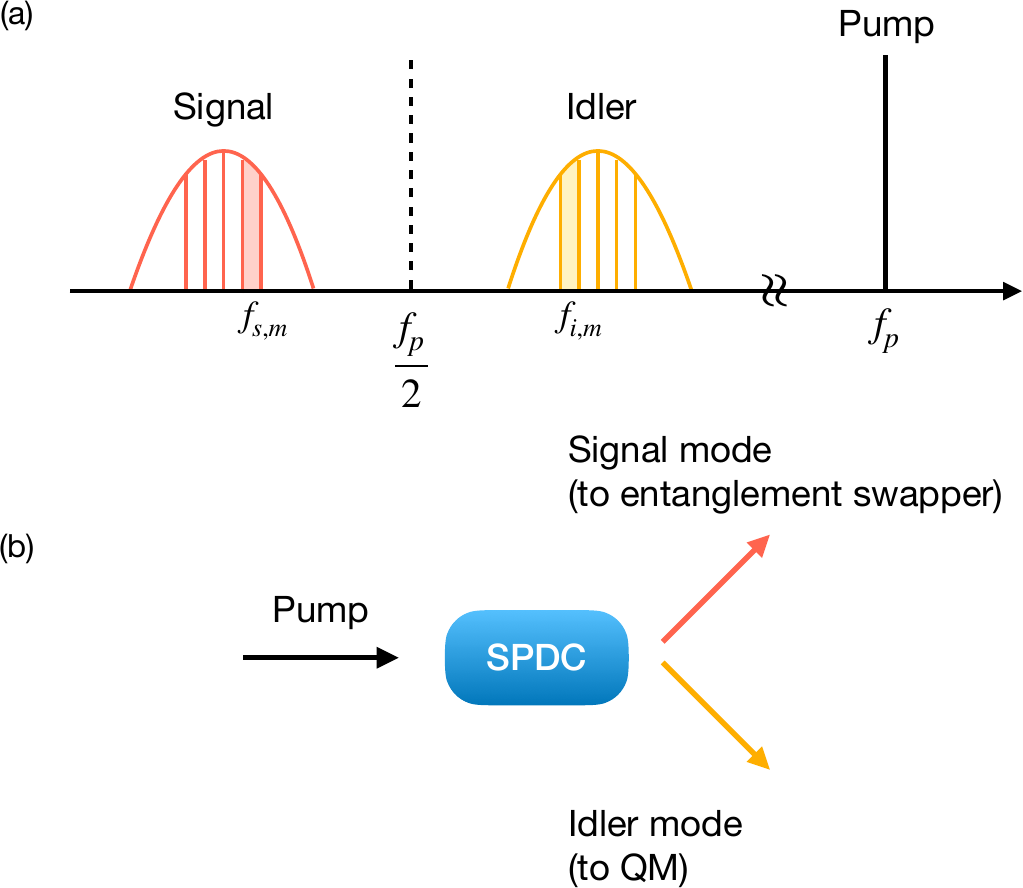}
\caption{Frequency multiplexed remote-entanglement generation. (a) Frequency spectrum of the pump, signal, and idler modes when frequency multiplexing is used. The shaded areas correspond to particular frequency bins for a signal-idler pair [indexed by $m$, symmetric about half of the pump frequency ($f_p/2$)]. The frequency of the signal mode is denoted by $f_{s,m}$ and the frequency of the idler mode is denoted by $f_{i,m}$. (b) Illustration of the SPDC emission modes and their respective destinations.}
\label{fig:freq_mul}
\end{figure}

The SPDC sources can generate frequency-multiplexed entangled photonic degrees of freedom in the Fock basis~\cite{grimau2017heralded}
\begin{align}
\ket{\text{SPDC}} &= \bigotimes_{m=1}^{N_\text{freq}}\sqrt{1-\lambda^2}\sum_{n=0}^{\infty}\lambda^n|n_{s,m}n_{i,m}\rangle,
\label{eqn:spdc_state}
\end{align}
where $N_\text{freq}$ is the number of frequency modes used for frequency multiplexing, $m$ is the index of a given signal-idler mode pair, $(s,m)$ refers to the signal mode, and $(i,m)$ refers to the idler mode for a specific mode pair [Fig.~\ref{fig:freq_mul}(a)].

In the actual calculation, we truncate the two-mode squeezed vacuum output state up to $\lambda^2$ terms. Note that a nondegenerate SPDC source can be used, allowing the signal mode to be at a telecommunication frequency and the idler mode at a QM-compatible frequency. Thus, the signal mode is directed to a remote-entanglement swapper, and the idler mode is stored in a local QM [Fig.~\ref{fig:freq_mul}(b)]. In this way, quantum frequency conversion is not necessary for the output photons~\cite{rielander2017frequency}. Once the heralding signals return from the remote-entanglement swapper, the information of a successfully entangled mode is known, which can be read out from the QM and directed to the next entanglement swapper in the case of a repeater chain.

However, a fundamental issue arises for scaling up the SPDC and QM-based repeater using the single-click entanglement swapping protocol, which is rooted in the multiphoton emission of the SPDC sources. The existence of multiphoton components and photon loss in the fiber transmission to the entanglement swappers makes the SPDC-based repeater design severely suffer from the false-heralding errors~\cite{hermans2023entangling}. This can be partially alleviated with a weak pump to avoid the multiphoton components. Nevertheless, it has been shown that when one optimizes the pump strength to avoid multiphoton emission, the achievable secret key rate for quantum key distribution is actually smaller than the direct transmission bound~\cite{guha2015rate}. This significantly limits the scalability of the SPDC-based repeater chains for the use of quantum information processing tasks. Moreover, in this repeater design, the elementary-link entanglement is encoded in an optical degree of freedom, and hence it is difficult to further process it deterministically, as is necessary in the case of executing entanglement distillation protocols. Together with the probabilistic entanglement swapping at the remote-entanglement swapper to merge the elementary links, achieving high-fidelity entanglement with the SPDC-based quantum repeater poses a significant challenge.

\section{Hybrid quantum repeaters}\label{sec:architecture}

In this section, we present the design of our hybrid repeater and the protocol for it to establish long-range entanglement. Recognizing the challenges mentioned in the preceding section, in our hybrid repeater proposal, we use SPDC sources as fast sources for remote-entanglement generation with heavy multiplexing. After the remote entanglement is heralded, we choose to transfer the entanglement to atomic emitters such that deterministic quantum operations can be performed. This avoids the error propagation in the case of an SPDC-based repeater.

The end-to-end entangled state we are aiming to establish is the Bell state between atomic qubits in the QPUs [Eq.~\eqref{eqn:bell_state}]. In the hardware design, we consider using cavity-coupled ion traps~\cite{takahashi2020strong} as the QPUs; however, the same design can also be applied to neutral-atom-based quantum devices, which are also known to be good platforms for atom-photon interface~\cite{li2024high,sinclair2025fault}.

\subsection{Hardware architecture}\label{subsec:hybrid}

\begin{figure*}[t]
\centering
\includegraphics[width=0.9\linewidth]{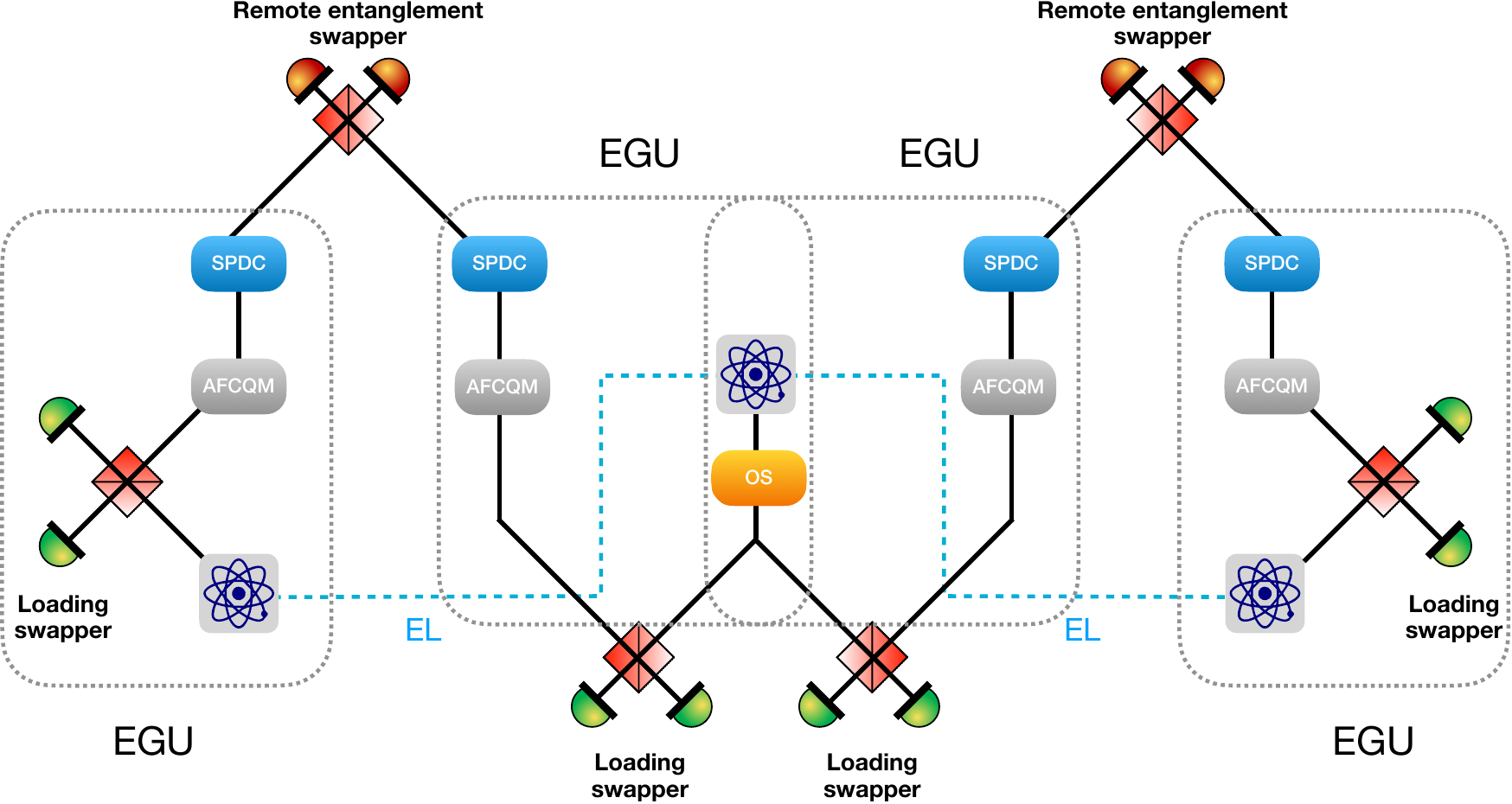}
\caption{Architecture of the hybrid quantum repeater (a one-hop chain). A one-hop linear hybrid quantum repeater chain is illustrated. Each entanglement generation unit comprises an SPDC source, an AFCQM, a loading swapper, and an atom-based QPU. The SPDC sources and remote-entanglement swappers generate the memory-memory entanglement heralded by a single-photon detection at the remote-entanglement swapper. The entanglement is then loaded into QPUs with additional local entanglement swapping and thus can be processed with deterministic quantum operations. The intermediate QPU has an additional optical switch (OS) such that the intermediate QPU can generate entanglement with both sides. The blue dashed lines indicate the elementary links of the hybrid repeater chain. A one-hop chain contains one repeater station between the end nodes.}
\label{fig:hybrid_repeater}
\end{figure*}

The proposed hybrid repeater architecture is illustrated in Fig.~\ref{fig:hybrid_repeater} (a one-hop linear hybrid repeater chain). An elementary link of the hybrid quantum repeater comprises two entanglement generation units (EGUs) and a remote-entanglement swapper. Each EGU comprises an SPDC source, an AFCQM, an atom-based QPU, and a local entanglement swapper, which we refer to as the loading swapper.

The establishment of ELs begins with the generation of photonic entanglement using SPDC sources [Eq.~\eqref{eqn:spdc_state}]. One half of the SPDC outputs (idler modes) are stored in the QMs and the other half (signal modes) are sent to the remote-entanglement swapper. The remote swapping generates entanglement between the QMs. This step is done with heavy spectrotemporal multiplexing due to the multimode capability of the AFCQM~\cite{bonarota2011highly,wei2024quantum}. After the QM-QM entanglement is generated, it is loaded to the QPUs via the loading swappers. Loading the entanglement to the QPUs enables subsequent deterministic quantum operations on the raw entangled pairs. After the loading is completed, the entanglement distillation and merging of elementary links are done with the QPUs.

In our hybrid architecture, high-visibility interference between SPDC emission and cavity-coupled atomic QPU emission is essential for the successful loading of entanglement. We note that in standard free-space SPDC configurations, the spectral bandwidths of these two sources are vastly different: The former typically resides in the terahertz regime~\cite{england2015storage}, while the latter is usually in the megahertz range~\cite{keller2004continuous}. This discrepancy results in a massive temporal mismatch that fundamentally limits the indistinguishability required for direct interference. Nevertheless, we expect that cavity-enhanced SPDC sources~\cite{bao2008generation} and advanced pulse-shaping techniques~\cite{ansari2018heralded} can be employed to bridge this gap, ensuring the high-visibility interference required for loading of the entanglement.

Next we provide the detailed protocol for generating long-range entanglement with the hybrid repeater.

\subsection{Protocol for remote-entanglement distribution using the hybrid repeater}

Here we present the step-by-step protocol for establishing end-to-end entanglement using our hybrid repeater. For each step, we provide the ideal quantum state resulting from that operation. The explicit derivation of these state transformations is detailed in Appendix~\ref{sec:appendix_state_change}. The first stage of the entanglement generation begins by pumping the SPDC sources to obtain two-mode entangled states in the Fock basis. We assume that the SPDC sources are excited by a pulsed laser, and the separation of the excitation pulses defines the temporal modes with $N_\text{temp}$ modes in total. In addition to the temporal multiplexing, we assume that a pump with a suitable linewidth is chosen such that the SPDC sources also operate with frequency multiplexing with $N_\text{freq}$ frequency modes. The state of a temporal mode is modeled by Eq.~\eqref{eqn:spdc_state}, and the joint state of two SPDC sources in this step is
\begin{align}
\ket{\text{SPDC}}^{\otimes 2} \propto \bigotimes^{N_\text{freq}}_{m=1}\Big(
&|0_{s_1,m}0_{i_1,m}0_{s_2,m}0_{i_2,m}\rangle \nonumber\\
+\lambda &|1_{s_1,m}1_{i_1,m}0_{s_2,m}0_{i_2,m}\rangle \nonumber\\
+\lambda &|0_{s_1,m}0_{i_1,m}1_{s_2,m}1_{i_2,m}\rangle \nonumber\\
+\lambda^2 &|2_{s_1,m}2_{i_1,m}0_{s_2,m}0_{i_2,m}\rangle \nonumber\\
+\lambda^2 &|0_{s_1,m}0_{i_1,m}2_{s_2,m}2_{i_2,m}\rangle \nonumber\\
+\lambda^2 &|1_{s_1,m}1_{i_1,m}1_{s_2,m}1_{i_2,m}\rangle\Big),
\label{eqn:joint_SPDC}
\end{align}
where the number subscripts $\{1,2\}$ correspond to the left EGU and right EGU, respectively.

After the generation of multiplexed entangled photonic states, the QM-compatible idler modes are stored in the local QM with the AFC protocol. The other telecom-frequency signal modes are sent to a remote-entanglement swapper, which is able to perform frequency-resolved entanglement swapping~\cite{merkouche2022heralding}. The single-click entanglement swapping protocol is executed at the remote-entanglement swapper to generate the QM-QM entanglement. Therefore, a single click in one of the photon detectors heralds the generation of remote entanglement at a particular temporal-frequency mode. We assume a successful heralding event is recorded for a particular temporal-frequency mode. We proceed with the heralded mode and omit the frequency mode index $m$ for brevity. We define the QM-QM entanglement established in this stage as
\begin{align}
|\psi_\text{QM}\rangle \propto |0_\text{QM1}1_\text{QM2}\rangle + |1_\text{QM1}0_\text{QM2}\rangle,
\end{align}
where the subscripts QM1 and QM2 correspond to the QM state of the left and right EGUs, respectively.

Upon receiving heralding signals from the remote-entanglement swapper, we proceed to load the entanglement. We employ local entanglement swapping to map the QM-QM entanglement onto the QPUs.

The QPUs generate atom-photon entanglement as described in Eq.~\eqref{eqn:ion_photon}. The emitted photons undergo frequency conversion to match the photons retrieved from the QMs. The converted photons then interfere with the memory output. If both local entanglement swappers record a single-click event, the resulting QPU-QPU entangled state is
\begin{align}
|\psi_\text{QPU-QPU}\rangle = \sqrt{\frac{1}{2}}\left(|D_1 B_2\rangle + |B_1 D_2\rangle\right).
\end{align}

Note that the loading steps for the two remote QPUs do not have to be done at the same time. This alleviates the need for remote synchronization in the loading stage.

Once the entanglement is loaded to the QPUs, the QPUs execute deterministic gate operations to perform entanglement distillation before merging different elementary links. The details regarding entanglement distillation and other error-suppression methods are discussed in Sec.~\ref{sec:error_mitigation}.

As a final step, the distilled elementary links stored in the QPUs are connected via deterministic Bell-state measurements to establish long-range entanglement. Note that the outcomes of these measurements must be fed forward to the end nodes to perform the necessary Pauli-frame correction on the final end-to-end entangled pair.

\subsection{Advantages and disadvantages of the hybrid design}

Compared to the designs of atom-based repeaters and SPDC-based repeaters, the proposed hybrid repeater incorporates devices with different strengths to establish long-range entanglement. Unlike the case of a QPU-based repeater, where the multiplexing capability is limited by the number of emitters in the QPU, the capacity of the AFCQM is determined by the inhomogeneous broadening of the rare-earth-doped material. Storage of more than 1000 modes by spectrotemporal multiplexing is possible with current technologies~\cite{bonarota2011highly,wei2024quantum}. The multiplexing increases the rate of generating elementary links.

Furthermore, compared to the SPDC-based repeater, the hybrid quantum repeater offers the ability to perform deterministic quantum operations. This enables the execution of entanglement distillation protocols and deterministic merging of elementary links. Advanced entanglement distillation protocols can be employed to suppress various quantum noises.

On the other hand, although the heavy multiplexing promises a faster rate for elementary-link generation, the usage of additional probabilistic entanglement swapping for loading the entanglement inevitably reduces the success probability of generating the elementary links.

The usage of additional components could also lead to unsatisfactory fidelity. Specifically, we note that for one generation of the elementary link, three swaps are needed. The false-heralding signals in each entanglement swapper can degrade the fidelity of the elementary links.

Finally, we employ a single-click entanglement swapping protocol due to its higher success probability than double-click protocols. As a result, this imposes a stringent requirement for phase stabilization~\cite{hermans2023entangling}. In Sec.~\ref{sec:error_mitigation} we discuss a specific entanglement distillation protocol that can alleviate the need for precise phase locking.

\section{Error-suppression strategy}
\label{sec:error_mitigation}
As discussed in the preceding section, the hybrid design potentially achieves superior entanglement distribution rates due to its extensive multiplexing capability and the capability for entanglement distillation. However, the use of more devices in a single repeater unit inevitably creates more avenues for errors. In this section, we investigate three suitable error-suppression strategies for the proposed hybrid repeater, focusing on photon-loss errors. Specifically, we consider three main loss channels: (i) transmission loss from the EGU to the remote-entanglement swapper, (ii) QM readout loss, and (iii) quantum frequency conversion loss for the atom-based QPU emission. We term (i) \emph{remote loss} and group both (ii) and (iii) as \emph{local loss}. In contrast to these loss channels, for the entanglement swapping, we assume the atom-photon entanglement is generated with unit fidelity, the photon detectors operate with unit efficiencies, and photon interference at the beam splitters occurs with perfect indistinguishability.

First, we calculate the density matrix for the elementary link (entanglement between QPUs) when one does not employ any error-suppression method (for general analytical forms of the density matrix for selected cases see Appendix~\ref{sec:appendix_dm}). For the following numerical examples, we choose the brightness parameters $\lambda = 0.1$ and $q=0.1$ for the SPDC source [see Eq.~(\ref{eqn:spdc_state})] and for the atomic emitter [see Eq.~(\ref{eqn:ion_photon})], respectively. Furthermore, we assume the fiber transmission loss from the EGU to the remote-entanglement swapper to be $0.5$ (about 10 km for the telecommunication wavelength) and set the local loss to zero. In Sec.~\ref{sec:error_supression_protocols} we investigate the effect of finite local loss.

The density matrix and the fidelity of the raw elementary link with the above parameters are
\begin{align}
	\rho_\text{raw} &=\bordermatrix{
			    ~ & BB & BD & DB & DD \cr
		BB & 0.094 & 0 & 0 & 0 \cr
    BD & 0 & 0.425 & 0.422 & 0 \cr
		DB & 0 & 0.422 & 0.425 & 0 \cr
		DD & 0 & 0 & 0 & 0.056  } ,
		\nonumber \\
        F_\text{raw} &= 0.847.
        \label{eqn:EL_raw}
\end{align}
All density matrices in this paper are expressed in the basis defined by the row and column labels shown above. All values are rounded to three decimal places. The error in the raw elementary link stems from the false-heralding errors in the single-click protocol~\cite{hermans2023entangling}. The first source of false-heralding errors is \emph{photon bunching}. This refers to the case where two photons arrive at a non-number-resolving detector in the entanglement swapper. This results in a single click in the photon detector due to Hong-Ou-Mandel (HOM) interference~\cite{hong1987measurement}, whereas the quantum state of the emitters remains in a product state. The second source of false heralding is \emph{photon loss}. This refers to the case where both photon sources emit photons, but one of the photons is lost during the transmission. In this case, a single click is recorded at the entanglement swapper, but the which-path information is not erased at the entanglement swapper, still leading to a product state between the photon sources.

False-heralding errors can occur at both the remote-entanglement swapper and the local loading swapper, introducing $\ket{BB}\bra{BB}$ and $\ket{DD}\bra{DD}$ components into the density matrix. These errors also induce effective dephasing, as can be seen from the reduced off-diagonal terms. Notably, the $\ket{BB}\bra{BB}$ error component is slightly larger than $\ket{DD}\bra{DD}$ because HOM interference errors in the loading stage directly contribute to the $\ket{BB}\bra{BB}$ population. To suppress these errors, we propose the following strategies.

\subsection{Photon-number-resolving entanglement swapper}

Using photon-number-resolving (PNR) detectors is a straightforward way to detect the photon-bunching false-heralding error. In our design, there are two kinds of entanglement swappers, the remote-entanglement swapper and the loading swapper (see Fig.~\ref{fig:hybrid_repeater}). Employing PNR detectors at all entanglement swappers leads to the following improved state (with the same parameters as those for generating $\rho_\text{raw}$):
\begin{align}
    \rho_\text{PNR}&=
    \begin{pmatrix}
        0  & 0 & 0 & 0 \\
        0 & 0.478 & 0.478 & 0 \\
        0 & 0.478 & 0.478 & 0 \\
        0 & 0 & 0 & 0.043   
    \end{pmatrix}, \nonumber \\
    F_\text{PNR} &= 0.957.
    \label{eqn:EL_PNR}
\end{align}
For the elementary-link entanglement, using photon-number-resolving detectors eliminates the $\ket{BB}\bra{BB}$ error component. This is due to the fact that in the loading stage, photon bunching can be detected and discarded. However, false heralding because of photon loss cannot be eliminated. It is also noteworthy that a detector that can distinguish \emph{no photon}, \emph{one photon}, and \emph{many photons} is enough. Recently demonstrated PNR capabilities in low-photon numbers using superconducting nanowire single-photon detectors~\cite{cahall2017multi,kawasaki2022generation} are particularly suited for this purpose and are readily available in the laboratory~\cite{endo2021quantum}.

\subsection{Extreme photon-loss protocol}
\begin{figure}[tbp]
    \centering
\includegraphics[width=0.8\linewidth]{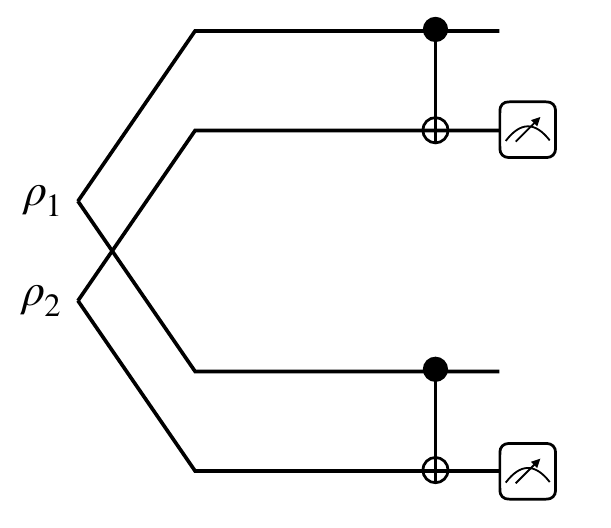}
\caption{Extreme photon-loss protocol. The EPL protocol takes two noisy entangled pairs, performs bilateral CNOT gates, and measures the target qubits. If the measurement outcomes are both $1$, the unmeasured pair is kept; otherwise, it is discarded.}
    \label{fig:EPL}
\end{figure}

The extreme photon-loss (EPL) protocol is an active entanglement distillation protocol that is specially designed for photon-loss error in the single-click entanglement swapping protocol~\cite{nickerson2014freely}. The quantum circuit for the protocol is depicted in Fig.~\ref{fig:EPL}. In the EPL protocol, two noisy entangled pairs are first shared. Then bilateral controlled-not (CNOT) gates are applied locally and the target qubits are measured. The unmeasured pair (control qubits) is kept only if the measurement outcomes are both $1$. The protocol is designed such that if the two entangled pairs are in the quantum state defined by the following equation, the EPL protocol leads to a perfectly entangled state $\ket{\Psi^+}$ upon success:
\begin{align}
    \rho_\text{noisy}= \paren{1-\alpha} \ket{\Psi^+} \bra{\Psi^+} + \alpha \ket{11}\bra{11} \label{eqn:EPL_state}
\end{align}
for $\alpha \in \paren{0,1}$. 

In our hybrid repeater, we choose $|B\rangle$ as the logical $|0\rangle$ and $|D\rangle$ as the logical $|1\rangle$ in the quantum circuit.
Application of the EPL protocol to $\rho_\text{raw}$ leads to 
\begin{align}
    \rho_\text{EPL}&=
    \begin{pmatrix}
        0.014  & 0 & 0 & 0 \\
        0 & 0.486 & 0.480 & 0 \\
        0 & 0.480 & 0.486 & 0 \\
        0 & 0 & 0 & 0.014
    \end{pmatrix}, \nonumber \\
    F_\text{EPL} &= 0.966. 
    \label{eqn:EL_EPL}
\end{align}
Compared to $\rho_\text{raw}$, both product-state error components are suppressed, but neither of them completely disappears. This can be understood as the EPL protocol can filter out the error where both input states are in $\ket{BB}\bra{BB}$ or $\ket{DD}\bra{DD}$. However, the case where one input state is in $\ket{BB}\bra{BB}$ and another input state is in $\ket{DD}\bra{DD}$ will pass the EPL protocol check. Therefore, the application of the EPL protocol to the raw elementary link actually symmetrizes the error component of $\ket{BB}\bra{BB}$ and $\ket{DD}\bra{DD}$ if both errors are present. We also note that, as mentioned in Ref.~\cite{nickerson2014freely}, the EPL protocol also alleviates the phase stabilization problem in the single-click entanglement swapping protocol. More specifically, if the two paths connecting to the entanglement swapper have a phase difference error, and if this error does not drift between two consecutive heralding events, the phase error can be eliminated by the EPL protocol.

\subsection{Reemission trick}

\begin{figure}[tbp]
    \centering
\includegraphics[width=1.\linewidth]{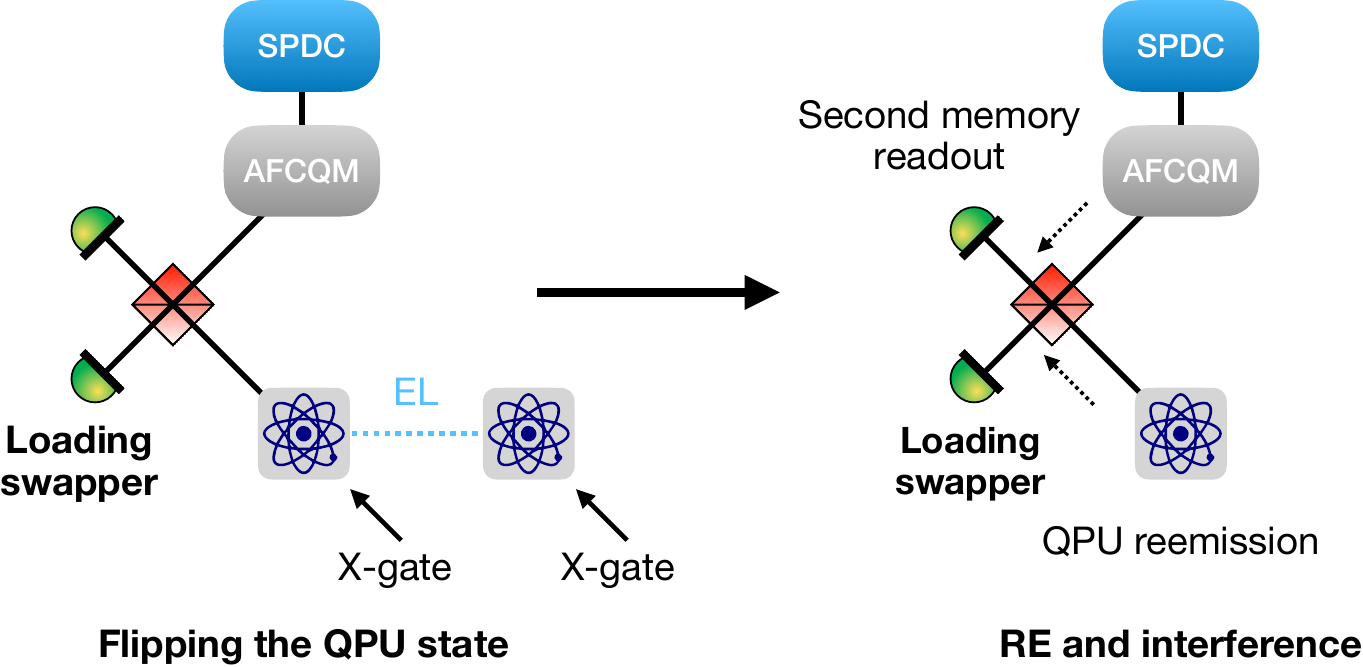}
    \caption{Reemission trick. The emitter (already entangled with another remote emitter) is first flipped and repumped to generate the state-dependent emission. The loading stage (interference with another output mode from the QM) is performed again.}
    \label{fig:BK_scheme}
\end{figure}

The EPL protocol requires active quantum gates to be performed directly on the two matter qubits in order to decrease the product-state ($\ket{BB}\bra{BB}$ or $\ket{DD}\bra{DD}$) error components. On the other hand, another error-suppression scheme exists that can achieve a similar effect but only needs local quantum operations performed on a single matter qubit. The reemission (RE) trick proposed by Barrett and Kok~\cite{barrett2005efficient} utilizes RE of photons and subsequent single-photon interference to achieve the error suppression. We modify the original procedure and show the resulting steps tailored to our hybrid repeater in Fig.~\ref{fig:BK_scheme}. After the first elementary link is generated, the atom emitters are first flipped by applying an $X$ gate ($\ket{B}\rightarrow\ket{D}$ and $\ket{D}\rightarrow\ket{B}$). This exchanges the error components of $\ket{BB}\bra{BB}$ and $\ket{DD}\bra{DD}$. The QPU generates the atom-photon entanglement again and interferes with another successfully entangled mode read out from the QMs. Note that the RE trick could, in theory, lead to more efficient qubit usage (for example, if the QPU only contains a single emitter for establishing one elementary link). The key insight for the RE trick is that if both emitters are in $\ket{DD}\bra{DD}$ after the bit flips, neither emits a photon during the RE stage. The reason for flipping the atomic emitter state is that the error component $\ket{BB}\bra{BB}$ has a higher value than that of $\ket{DD}\bra{DD}$ [Eq.~(\ref{eqn:EL_raw})]. Therefore, the error component $\ket{DD}\bra{DD}$ can be reduced by postselection. The elementary link with the RE trick error suppression is given by
    \begin{align}
    \rho_\text{RE}&=
    \begin{pmatrix}
        0.116  & 0 & 0 & 0 \\
        0 & 0.441 & 0.436 & 0 \\
        0 & 0.436 & 0.441 & 0 \\
        0 & 0 & 0 & 0.001
    \end{pmatrix}, \nonumber \\
    F_\text{RE} &= 0.877.
    \label{eqn:EL_RE}
\end{align}
As expected, the error component $\ket{DD}\bra{DD}$ is suppressed, but the component $\ket{BB}\bra{BB}$ is increased because of the false heralding due to photon bunching. The reason that the $\ket{DD}\bra{DD}$ is not completely suppressed is as follows. Even if both emitters are in a dark state, if both QMs emit photons (i.e., the memory was in the $|1_\text{QM1}1_\text{QM2}\rangle \langle 1_\text{QM1} 1_\text{QM2}|$ state), the loading swappers still record single-click events, but the atomic emitters remain in $|DD\rangle \langle DD|$.

\subsection{Error-suppression protocol for the hybrid repeater} \label{sec:error_supression_protocols}

As discussed in the preceding section, no single elementary strategy can eliminate all errors arising from photon loss in the hybrid repeater. We therefore combine multiple error-suppression methods to achieve improved performance. Moreover, the earlier analysis focused only on photon loss in the fiber transmission to the remote-entanglement swapper (remote loss). In this section we further consider the local loss and its impact on the error-suppression protocol.

\begin{figure}[tbp]
\centering
\includegraphics[width=\linewidth]{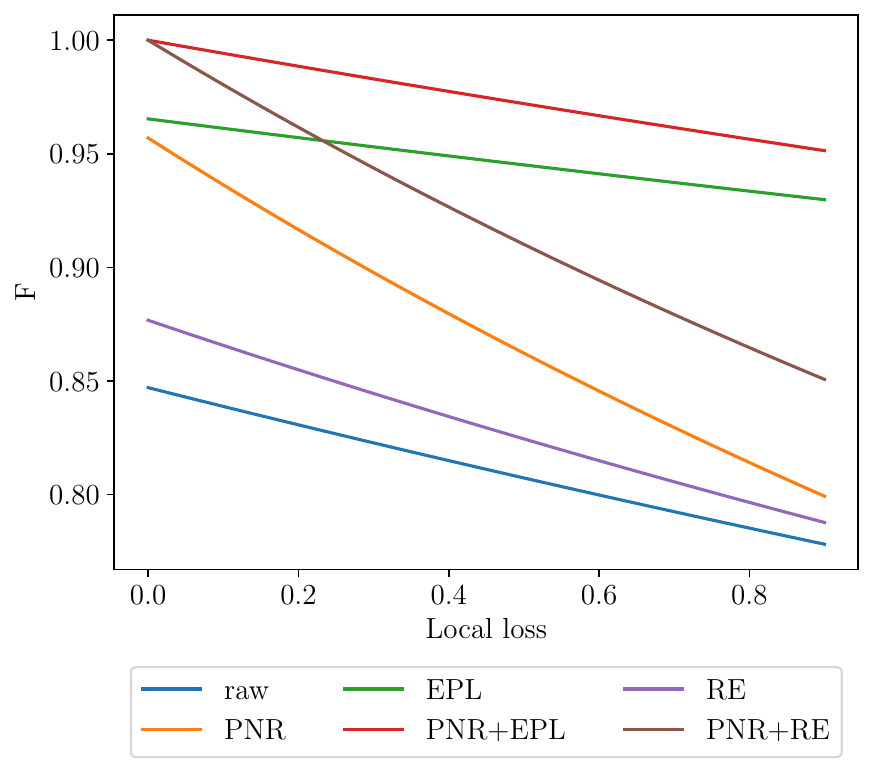}
\caption{Achievable fidelities under varying local-loss strengths for different error-suppression methods: raw, no error suppression; PNR, photon-number-resolving entanglement swapping; EPL, EPL entanglement distillation; PNR + EPL, EPL distillation combined with photon-number-resolving detection; RE, reemission trick; and PNR + RE, reemission trick combined with photon-number-resolving detection.}
\label{fig:fidelity_error_suppressions}
\end{figure}

In the following analysis, we assume that the two local-loss mechanisms (quantum frequency conversion and quantum-memory readout loss) share the same loss parameter. In Fig.~\ref{fig:fidelity_error_suppressions} we compare various error-suppression methods (including their combinations) and examine the achievable fidelities under finite local loss. We observe that the PNR + EPL strategy consistently achieves the highest fidelity across all local-loss regimes. We note that in the low-local-loss regimes, the PNR + RE strategy performs comparably to PNR + EPL. Nevertheless, in this work, we adopt the PNR + EPL method as the primary error-suppression strategy for our hybrid repeater protocol, as it enables simpler rate calculations and scheduling policies. A detailed analysis of RE-based error suppression is left for future work. We note that in the absence of local loss, both PNR + EPL and PNR + RE can achieve unit fidelity with respect to the $|\Psi^+\rangle$ Bell state.

The analytical expression for the distilled state $\rho_\text{PNR+EPL}$ is provided in Appendix~\ref{sec:appendix_dm}.

\section{Numerical results}\label{sec:results}

In this section, we investigate the performance of the hybrid repeater for remote-entanglement distribution. The baseline system we use as a benchmark is the atom-based repeater system (Fig.~\ref{fig:atom_based}). Specifically, we calculate the end-to-end entanglement-based quantum key distribution asymptotic key rate with the entanglement-based BBM92 protocol proposed by Bennett, Brassard, and Mermin~\cite{bennett1992quantum}. The secret key rate has the generic expression
\begin{align}
    R_\text{key} &= r_\text{rep} r_\text{sec}, \label{eqn:key-rate}
\end{align}
where $r_\text{rep}$ is the repetition rate for generating the end-to-end entangled pair and $r_\text{sec}$ the secret key fraction which depends on the end-to-end entangled state. For the secret key fraction, we use the expression~\cite{scarani2009security}
\begin{align}
    r_\text{sec} &= \max\{0, 1 - h(p_\text{bit}) - h(p_\text{phase})\},\label{eqn:secret_fraction}
\end{align}
where $p_\text{bit}$ ($p_\text{phase}$) is the quantum bit error rate when both end users perform measurements in the computational (diagonal) basis, and $h$ is the binary entropy function. We provide the entanglement distribution rates, i.e., the number of entangled pairs that can be shared from end to end per unit time in Appendix~\ref{sec:appendix_rate}.

In the following numerical simulations, the physical parameters listed in Table~\ref{tab:parameters} are used throughout. Apart from the parameters in Table~\ref{tab:parameters}, the atom-photon entanglement fidelity, photon detector efficiencies, and photon interference visibilities at the beam splitters are all assumed to be unity as in Sec.~\ref{sec:error_mitigation}. We apply PNR + EPL as the error-suppression strategy. Therefore, the use of PNR detectors with unit efficiency is assumed. Also, because of the EPL protocol to purify the elementary links, at least two qubits per QPU are necessary for generating one elementary link. The efficiencies for quantum frequency conversion and quantum memory readout are variables, and we consider a few representative cases in Secs.~\ref{subsec:multiplexing_brightness} and \ref{subsec:key_rate_comparison}.

\begin{table*}[t]
\caption{\label{tab:parameters}List of parameters used in the quantum repeater simulation. The parameter $N_\text{atom}=2$ represents the minimum number of atomic qubits required to implement the entanglement-distillation protocol. The QPU-related parameters are estimated based on values reported in the relevant literature~\cite{monz2009realization,bruzewicz2019trapped,myerson2008high}. For the atom-photon entanglement generation time, we incorporate the duration of atom-state preparation and assume that the emitted photon has a temporal width on the order of $1~\mu\mathrm{s}$, comparable to SPDC-based emission. The repetition time of the SPDC source photon generation, $t_\text{SPDC}$, is the interpulse spacing between successive temporal modes required to prevent temporal overlap.}
\centering
\small
\setlength{\tabcolsep}{4pt}
\begin{tabular}{lcr}
\hline\hline
Parameter & Description & Value \\
\hline
$c$ & Speed of light in fiber & $0.2~\mathrm{km}\,\mu\mathrm{s}^{-1}$ \\
$\alpha_\text{fiber}$ & Fiber attenuation coefficient & $0.3~\mathrm{dB}\,\mathrm{km}^{-1}$ \\
$\eta_\text{fiber}(d)$ & Fiber transmission efficiency over a distance $d$ & $=10^{-\alpha_\text{fiber}d/10}$ \\
$t_\text{atom}$ & \shortstack{Time for generating atom-photon entanglement, including\\ moving the atoms into the cavity} & $100~\mu\mathrm{s}$ \\
$t_\text{SPDC}$ & SPDC repetition time & $1~\mu\mathrm{s}$ \\
$t_\text{meas}$ & Time for measuring the atomic qubit & $100~\mu\mathrm{s}$ \\
$t_\text{CNOT}$ & Time for the controlled-not gate in the quantum processor & $100~\mu\mathrm{s}$ \\
$N_\text{atom}$ & Number of atomic qubits in the QPU used to generate an elementary link & $2$ \\
$N_\text{mul}$ & Total number of modes for multiplexing & $1000$ \\
\hline\hline
\end{tabular}
\end{table*}

In the following, we derive the analytical expressions for the repetition rate $r_\text{rep}$ for the atom-based and hybrid repeaters, respectively.

\subsection{Timing analysis and analytical repetition rate expressions}

\subsubsection{Atom-based repeaters}
For the atom-based repeaters, we assume temporal multiplexing is used, and there are $N_\text{QPU}$ atom-based QPUs in the repeater chain (hence $N_\text{QPU}-1$ elementary links). Each QPU contains $N_\text{atom}$ emitters to establish entanglement with one neighboring node. The time it takes for a single round of generating the remote entanglement comprises the following stages. The first stage is the attempt to generate elementary links with temporal multiplexing, which is done in parallel for every pair of QPUs. This stage takes time
\begin{align}
    t_\text{EL,try}^\text{atom}= N_\text{atom}t_\text{atom} + t_\text{signal},
\end{align}
where $t_\text{signal}$ is the time required for the photon to travel to the remote swapper plus the time for the heralding signal traveling back, i.e., $t_\text{signal} = \frac{d_\text{QPU}}{c}$, where $d_\text{QPU}$ is the distance between QPUs and $c$ is the speed of light in the fiber. We assume an equally spaced repeater chain. Therefore $d_\text{QPU}=\frac{D}{(N_\text{QPU}-1)}$, where $D$ is the end-to-end distance. After the elementary links are generated, the EPL entanglement distillation protocol is executed to distill the elementary links. The distillation process takes time
\begin{align}
    t_\text{ED} = t_\text{CNOT} + t_\text{meas} + t_\text{signal},\label{eqn:t_ED}
\end{align}
where $t_\text{CNOT}$ is the time for executing a CNOT gate in QPUs and $t_\text{meas}$ is the time for measuring the atom-based qubits. For every distillation attempt, the measurement outcome needs to be exchanged within an elementary link to determine if the distillation is successful. Hence, another $t_\text{signal}$ is consumed.
The total time an attempt takes to generate a distilled elementary link is
\begin{align}
    t_\text{EL,D}^\text{atom}= t_\text{EL,try}^\text{atom} + t_\text{ED}.
    \label{eqn:t_EL,d}
\end{align}
After all distilled elementary links have been generated, entanglement swapping is performed through deterministic Bell measurements in parallel at the intermediate QPUs. For classical feedforward, we conservatively allocate one full-chain propagation time, $(N_\text{QPU}-1)t_\text{signal}$, to communicate the swapping outcomes and update the endpoint Pauli frame. With this timing convention, the merge time is
\begin{align}
    t_\text{merge} &= t_\text{CNOT} + t_\text{meas} + (N_\text{QPU}-1)t_\text{signal},\nonumber \\
    &\qquad \forall\,N_\text{QPU}\geq 3.
    \label{eqn:t_merge}
\end{align}

Note that for the elementary link $N_\text{QPU}=2$, the merge is not required, and thus $t_\text{merge}=0$. Next, we discuss the success probability of each step. For each photon emitted from the QPU, we consider two loss channels, including quantum frequency conversion efficiency $\eta_\text{QFC}$ and fiber transmission efficiency $\eta_\text{fiber}(d)$, which depends on the photon traveling distance $d$. For brevity, we suppress the explicit dependence of $\eta_\text{fiber}(d)$ on the distance $d$ and write it as $\eta_\text{fiber}$ hereafter. Therefore, the total remote loss for a single arm from the QPU to the remote-entanglement swapper is
\begin{align}
    \varepsilon_r^\text{atom} = 1 - \eta_\text{QFC}\eta_\text{fiber},
\end{align}
and the click probability at the remote-entanglement swapper for each emitted photon $p_\text{click}^\text{atom}$ depends on the above remote loss. 

For the distillation of the elementary link, we assume the extreme photon-loss entanglement distillation protocol is used. Therefore, two entangled pairs between QPUs are necessary. Due to the temporal multiplexing, the success probability of generating at least two elementary links is given by
\begin{align}
    p_\text{EL,2}^\text{atom} = 1 &- \left(1-p_\text{click}^\text{atom}\right)^{N_\text{atom}} \nonumber \\ &- N_\text{atom}\left(p_\text{click}^\text{atom}\right) \left(1-p_\text{click}^\text{atom}\right)^{N_\text{atom}-1},
\end{align}
where the first subtracted term corresponds to the probability that no entangled pair is generated and the second subtracted term corresponds to the probability that only one entangled pair is generated.

Let the probability of successfully executing the extreme photon-loss entanglement distillation protocol be $p_\text{ED}$. The probability of generating a distilled elementary link between two QPUs is
\begin{align}
    p_\text{EL,D}^\text{atom} = p_\text{EL,2}^\text{atom} p_\text{ED}.
\end{align}
In this paper, we assume a simplified fixed-slot scheduling policy; for each distilled elementary link generation, if any step fails, the elementary link is regenerated. The expected time to generate distilled elementary links (in parallel) between every pair of QPUs is given by the expression (derivation in Appendix~\ref{sec:appendix_t_EL_D_all})
\begin{align}
    \langle t_\text{EL,D,all}^\text{atom} \rangle  
    &= t_\text{EL,D}^\text{atom} \sum^{N_\text{QPU}-1}_{j=1} \left(-1\right)^{j+1} \binom{N_\text{QPU}-1}{j} \nonumber \\ &\times \frac{1}{1-\left(1-p_\text{EL,D}^\text{atom}\right)^j}
    \label{eqn:t_EL_d_all}.
\end{align}
Finally, the expected time to generate end-to-end entanglement is
\begin{align}
    \langle t_\text{end}^\text{atom}\rangle = \langle t_\text{EL,D,all}^\text{atom}  \rangle + t_\text{merge}.\label{eqn:t_end}
\end{align}
The overall repetition rate is defined as the inverse of the expected time
\begin{align}
    r_\text{rep}^\text{atom} = \frac{1}{\langle t_\text{end}^\text{atom} \rangle}.\label{eqn:R_end}
\end{align}

\subsubsection{Hybrid repeaters}
For the hybrid repeater, we assume $N_\text{freq}$ frequency modes and $N_\text{temp}$ temporal modes for the SPDC are used for spectrotemporal multiplexing, and the remote-entanglement swapper has the ability to perform frequency-resolving single-photon detection~\cite{merkouche2022heralding}. As we described in the hybrid protocol, we also need two entangled pairs between the QPUs for the EPL entanglement distillation.

In the first phase of the hybrid repeater protocol, the SPDCs are pumped to generate an entangled photonic state in the Fock basis [Eq.~(\ref{eqn:spdc_state}), repeated $N_\text{temp}$ times]. The remote-entanglement swapper then sends the information on which frequency mode has been successfully detected back to the QMs. This step takes time
\begin{align}
    t_\text{QM,try} = N_\text{temp}t_\text{SPDC} + t_\text{signal}.
\end{align}
In the second phase, the generation of QPU entanglement is attempted by loading the photonic entanglement. The QMs then read out the heralded modes sequentially. The time of this phase depends on the QPU emission speed as well as the number of temporal modes. Note that for each loading attempt, the QPU stations need to communicate to make sure both loading swappers have a single-click heralding signal. If one loading attempt is not successful, QPUs can retry loading until all the QM temporal modes are exhausted. Since the loading time is upper bounded by trying to load all available temporal modes, we take the loading time
\begin{align}
    t_\text{load} = N_\text{temp} \left( t_\text{atom} + t_\text{signal} \right).
\end{align}
Therefore, the total time required for one QPU-QPU entanglement generation trial is
\begin{align}
    t_\text{EL,try}^\text{hybrid} = t_\text{QM,try} + t_\text{load}.
\end{align}
Once the loading of two entangled pairs is finished, the EPL entanglement distillation protocol is performed. The total duration of one attempt to generate a distilled EL is therefore
\begin{align}
    t_\text{EL,D}^\text{hybrid} = t_\text{EL,try}^\text{hybrid} + t_\text{ED}.
\end{align}
If the QPU-QPU entanglement is generated across every pair of QPUs, then the same $t_\text{merge}$ as Eq.~(\ref{eqn:t_merge}) is consumed to establish the end-to-end entanglement.

Next we analyze the success probabilities for each step. Since the signal mode sent to the remote-entanglement swapper is assumed to be at the telecommunication frequency, the remote loss from the SPDC source to the remote-entanglement swapper is
\begin{align}
    \varepsilon_r^\text{hybrid} = 1 - \eta_\text{fiber}.
\end{align}
Let the click probability for heralding the generation of QM-QM entanglement for each frequency mode per temporal mode be $p_\text{click,freq}^\text{hybrid}$. Since frequency multiplexing is employed at each emission, the probability of heralding a pair generation for a temporal mode in at least one of the frequency modes is
\begin{align}
p_\text{click,temp}^\text{hybrid}= 1 - \left(1-p_\text{click,freq}^\text{hybrid}\right)^{N_\text{freq}}.
\end{align}
For simplicity, if multiple frequency modes are successfully heralded within the same temporal mode, we select only one successful mode for subsequent loading. Next, the QPU is programmed to generate an atom-photon entangled state where the photon undergoes the quantum frequency conversion loss channel, parametrized by $\eta_\text{QFC}$. On the other hand, the photon output from the QM undergoes a loss channel parametrized by $\eta_\text{QM}$. Since the loading is performed locally, we assume no fiber loss.

The loading stage has overall click probability $p_\text{load}$ which depends on $\eta_\text{QFC}$ and $\eta_\text{QM}$.  For each temporal mode, the probability to produce one elementary link is
\begin{align}
    p_\text{EL}^\text{hybrid} = p_\text{click,temp}^\text{hybrid} p_\text{load}.
\end{align}
Since there are $N_\text{temp}$ temporal modes, the overall probability of generating at least two elementary links is 
\begin{align}
    p_\text{EL,2}^\text{hybrid}&=1 - \left(1 - p_\text{EL}^\text{hybrid}\right)^{N_\text{temp}} \nonumber \\
    &- N_\text{temp} p_\text{EL}^\text{hybrid} \left(1-p_\text{EL}^\text{hybrid}\right)^{N_\text{temp}-1}.
\end{align}
The success probability of the entanglement distillation is also $p_\text{ED}$ as defined previously. Therefore, the probability of generating a distilled elementary link is
\begin{align}
    p_\text{EL,D}^\text{hybrid} = p_\text{EL,2}^\text{hybrid} p_\text{ED}.
\end{align}
The expected time to generate distilled elementary links between every pair of QPUs has the form (derivation in Appendix~\ref{sec:appendix_t_EL_D_all})
\begin{align}
    \langle t_\text{EL,D,all}^\text{hybrid} \rangle
    &= t_\text{EL,D}^\text{hybrid} \sum^{N_\text{QPU}-1}_{j=1} \left(-1\right)^{j+1} \binom{N_\text{QPU}-1}{j} \nonumber \\ &\times \frac{1}{1-\left(1-p_\text{EL,D}^\text{hybrid}\right)^j}
\end{align}
and the expected time to generate end-to-end entanglement is
\begin{align}
    \langle t_\text{end}^\text{hybrid}\rangle = \langle t_\text{EL,D,all}^\text{hybrid}  \rangle + t_\text{merge}.
\end{align}
This leads to the overall rate expression
\begin{align}
    r_\text{rep}^\text{hybrid} = \frac{1}{\langle t_\text{end}^\text{hybrid} \rangle}.
\end{align}

\subsection{Multiplexing and brightness optimization}\label{subsec:multiplexing_brightness}
In our proposal, we employ both temporal multiplexing $N_\text{temp}$ and frequency multiplexing $N_\text{freq}$. Although it is evident that the end-to-end entanglement-generation rate benefits from multiplexing, the optimal distribution of the total number of multiplexed modes $N_\text{mul}$ to $N_\text{temp}$ and $N_\text{freq}$, with $N_\text{mul}=N_
\text{temp} N_\text{freq}$, is unclear. We refer to this optimization task as the \emph{multiplexing partition problem}.

\begin{figure*}[t]
    \centering
    \includegraphics[width=1\linewidth]{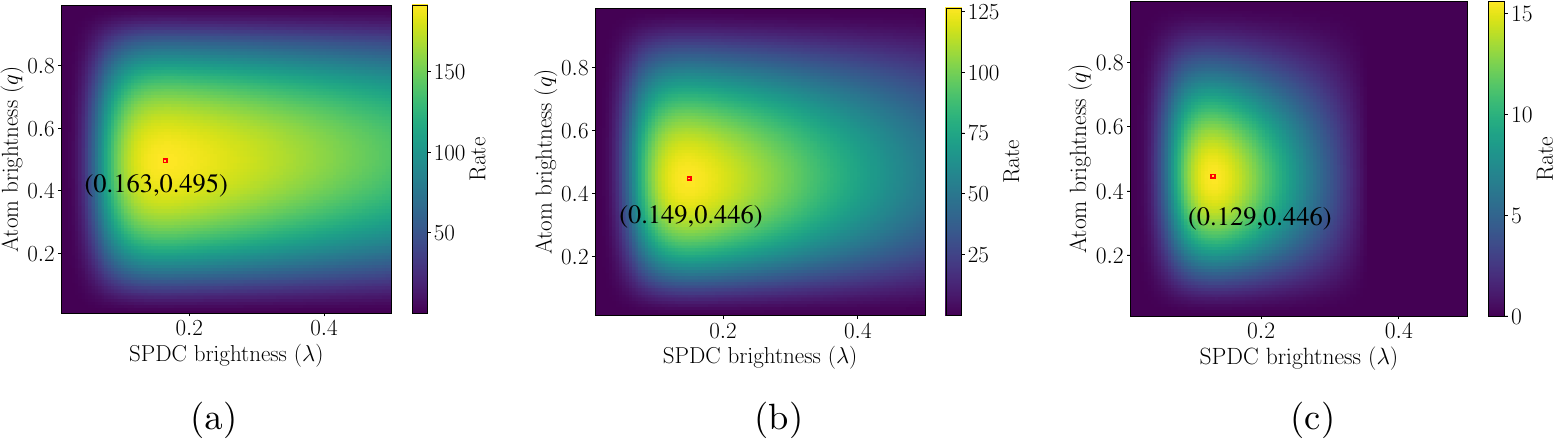}
    \caption{Brightness optimization. The colormaps correspond to the secret key rates with different brightness parameters for different local-loss parameters (a) $\eta_\text{QFC}=0.99$ and $\eta_\text{QM}=0.99$, (b) $\eta_\text{QFC}=0.9$ and $\eta_\text{QM}=0.9$, and (c) $\eta_\text{QFC}=0.5$ and $\eta_\text{QM}=0.5$. At a given distance ($d=10$ km and $N_\text{QPU}=2$ in this example), both SPDC brightness $\lambda$ and atom brightness $q$ are optimized to obtain the optimal key rates. The other parameters used in the simulation are shown in Table~\ref{tab:parameters}. In the case of low local loss, the optimal atom brightness is close to $0.5$. In the case of large local loss, the optimal atom brightness decreases due to the false-heralding error in the loading stage.}
    \label{fig:brightness}
\end{figure*}

On the one hand, increasing $N_\text{freq}$ improves the per-temporal-mode success probability for generating remote entanglement. On the other hand, it reduces the number of available loading attempts. The reverse tradeoff holds when $N_\text{temp}$ is increased. To determine the optimal multiplexing partition, we perform a brute-force search over all possible allocations of temporal and frequency modes, assuming a total multiplexing budget of $N_\text{mul} = 1000$. In particular, we compute the secret key rates for a single elementary link of the hybrid repeater as a function of $N_\text{temp}$ and distance, with the corresponding $N_\text{freq} = N_\text{mul} / N_\text{temp}$. Furthermore, at each given distance, the optimal brightness parameters [$q$ in Eq.~(\ref{eqn:ion_photon}) and $\lambda$ in Eq.~(\ref{eqn:spdc_state})] are obtained via direct numerical optimization performed in $0.01\leq q\leq 0.99$ and $0.01\leq\lambda\leq 0.5$. Figure~\ref{fig:brightness} shows secret key rates at $d=10$ km as a function of the brightness parameters for three sets of local losses, from which optimal brightness parameters can be extracted.

\begin{figure}[tbp]
    \centering
    \includegraphics[width=1\linewidth]{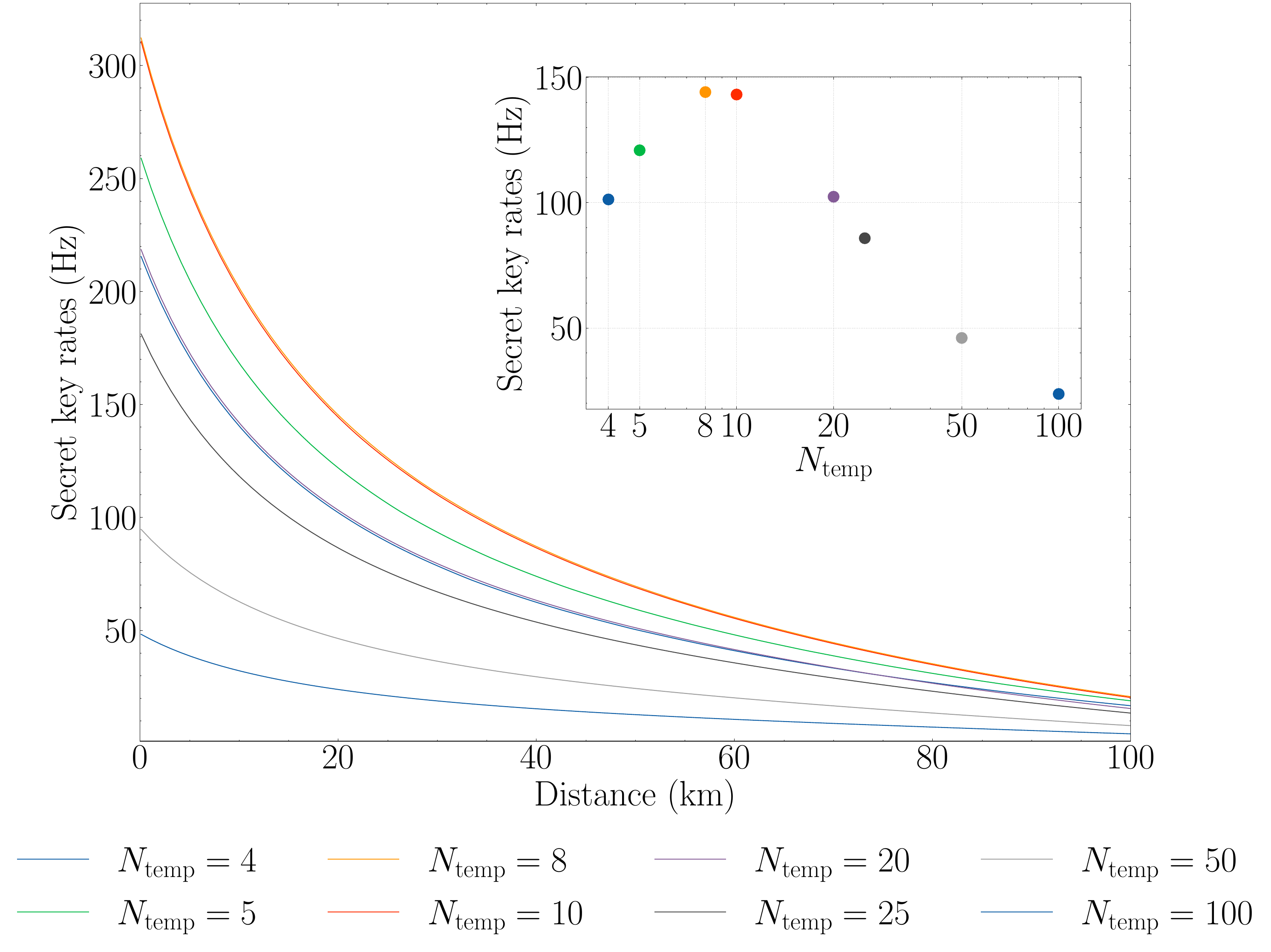}
    \caption{Multiplexing partition tradeoff. The achievable key rates for an elementary link of the hybrid repeater with different numbers of temporal modes are shown. Out of all multiplexing partitions, the $N_\text{temp}=8$ and $N_\text{freq}=125$ partition achieves the highest rates. The inset shows the secret key rates versus $N_\text{temp}$ at $d=20$ km.}
    \label{fig:multiplexing_tradeoff}
\end{figure}

The numerical results for representative multiplexing partitions are presented in Fig.~\ref{fig:multiplexing_tradeoff} (assuming zero local loss). For $4\leq N_\text{temp}\leq 8$, the secret key rate increases with $N_\text{temp}$, indicating that the benefit of additional loading attempts outweighs the reduced frequency multiplexing. However, when $N_\text{temp}>8$, the loss of frequency multiplexing reduces the remote-entanglement generation probability and increases the duration of the loading stage, both of which negatively impact the overall key rate. Based on these results, for a total multiplexing capacity of $N_\text{mul}=1000$, we choose $N_\text{temp}=8$ and $N_\text{freq}=125$ in all subsequent numerical calculations regarding hybrid repeaters. General methods for optimizing multiplexing partitions under realistic noise are left for future research.

\subsection{Secret key rate comparison}\label{subsec:key_rate_comparison}
To investigate the performance of our hybrid repeater for long-distance quantum communication, we numerically evaluate the end-to-end secret key rate using a linear repeater-chain model. As a baseline for comparison, we consider atom-based repeater chains. For a given distance, we examine up to two hops in the repeater chain, where zero hops corresponds to a single elementary link and two hops corresponds to two intermediate repeater stations. We explicitly simulate the state transformation at each entanglement-swapping step as well as the merging of elementary links.

\begin{figure*}[t]
\centering
\includegraphics[width=1\linewidth]{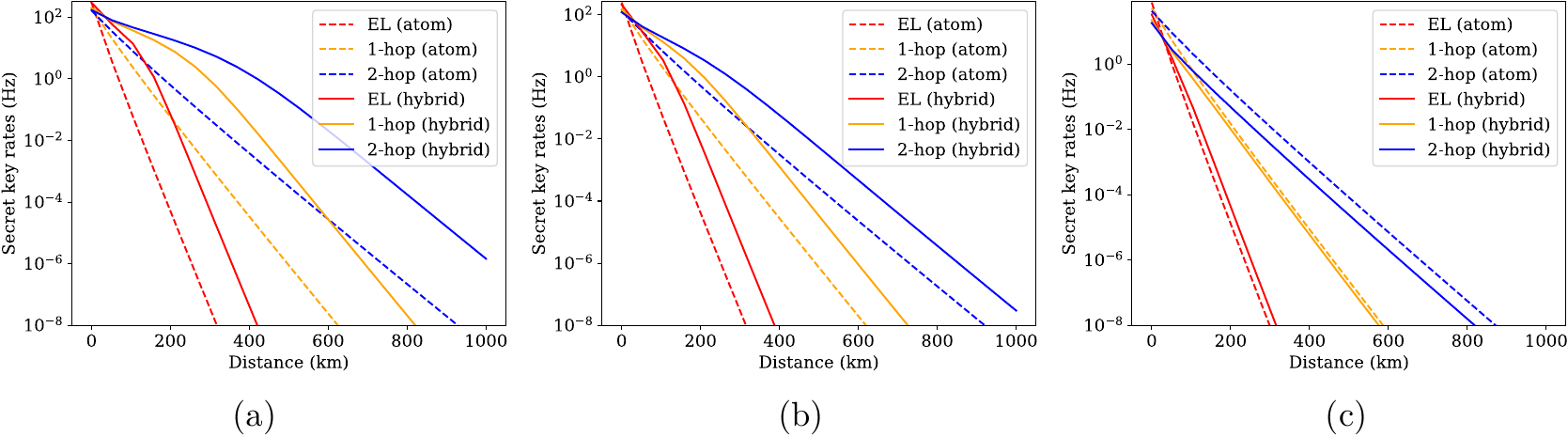}
\caption{End-to-end secret key rates: achievable secret key rates as a function of end-to-end distance. Dashed lines indicate the performance of atom-based repeater chains, while solid lines show the performance of the hybrid repeater. We evaluate three parameter regimes: (a) an idealized scenario representative of future devices ($\eta_\text{QFC}=0.99$ and $\eta_\text{QM}=0.99$), (b) a near-term optimistic regime ($\eta_\text{QFC}=0.9$ and $\eta_\text{QM}=0.9$), and (c) a regime based on conservative estimation of current experimental parameters ($\eta_\text{QFC}=0.5$ and $\eta_\text{QM}=0.5$).}
\label{fig:key_rates}
\end{figure*}

Figure~\ref{fig:key_rates} shows the resulting key rates as a function of distance for different hop numbers. At each distance, the brightness of the emitter is optimized to obtain the optimal secret key rate. We present results for three sets of local-loss parameters: an idealized scenario [Fig.~\ref{fig:key_rates}(a)], a near-term optimistic scenario [Fig.~\ref{fig:key_rates}(b)], and a regime consistent with current device performance [Fig.~\ref{fig:key_rates}(c)]. In both Figs.~\ref{fig:key_rates}(a) and \ref{fig:key_rates}(b), the hybrid repeater achieves higher key rates than the atom-based architecture across all distances, despite the extra overhead introduced by the photonic-to-atomic loading step.

A notable qualitative difference appears in the rate-distance scaling. For atom-based repeater chains, the key rate decays exponentially with distance for any fixed number of hops. In contrast, each solid curve for the hybrid repeater exhibits two distinct scaling regimes. In the short-distance regime, the decreasing trend is significantly slower than in the purely atom-based case. This behavior arises because remote entanglement between QMs can be generated almost deterministically; the rate is instead limited by the decreasing loading probability and the entanglement-distillation overhead as the distance increases. In the long-distance regime, the scaling of the hybrid repeater eventually matches that of the atom-based repeater, but with a constant multiplicative advantage due to spectrotemporal multiplexing.

In the current-technology regime shown in Fig.~\ref{fig:key_rates}(c), the advantage of the hybrid repeater over the atom-based counterpart appears only at the elementary-link connection stage. This behavior arises because, under large local losses, the residual noise from elementary-link generation cannot be fully removed by elementary-link-level entanglement distillation. These imperfections persist in the Bell-state output from the elementary link and propagate through the deterministic entanglement-merging stages. As more hops are added, this accumulated noise is amplified along the chain of QPUs. Consequently, although the hybrid scheme retains an advantage in the first connection step, its key rate deteriorates much more rapidly than in low- or moderate-loss regimes, severely compromising its ability to support genuinely long-distance quantum communication.

This problem may be alleviated by improving local efficiencies ($\eta_\text{QFC}$ and $\eta_\text{QM}$) or by employing nested entanglement distillation~\cite{dur1999quantum}, where distillation is performed not only on elementary links but also after each merging step to suppress accumulated errors. Determining the optimal nested distillation strategy and corresponding network scheduling policies is a direction for future research.

\section{Discussion}\label{sec:discussion}

We have proposed an efficient hybrid quantum-repeater architecture that integrates photonic components with heavy temporal and spectral multiplexing enabled by AFCQMs. In parallel, we employed a simple yet effective method for engineering and distilling entanglement within the repeater chain, explicitly addressing the challenge posed by multiphoton events in SPDC-based sources through a tailored error-suppression strategy. Note that our proposal can be applied not only to long-distance networks but also to a local cluster of interconnected QPUs to perform distributed quantum computing, in particular when the interconnecting optical channels have non-negligible losses.

 We have identified a clear quantitative advantage of the hybrid repeater over the atom-based repeater in the optimistic parameter regimes. However, our analysis also highlights that this advantage is sensitive to local losses and that improving the efficiency of local quantum operations is crucial for scaling the proposed design to many hops.

A key advantage of our design is the substantial enhancement in entanglement-distribution rate enabled by heavy spectrotemporal multiplexing during the QM-QM entanglement-generation stage. This capability effectively mitigates the probabilistic nature of photonic entanglement generation and brings the architecture within the realistic reach of near-term experimental technology.

Our work also points toward several promising directions for further development. First, the present analysis employs a simplified noise model capturing only the dominant error mechanisms. Incorporating more detailed device-level imperfections, such as finite memory lifetimes, crosstalk between frequency modes, and gate errors in QPUs, would allow for more accurate performance estimates and hardware-tailored distillation protocols. Second, our study focuses on linear repeater chains distributing Bell pairs. Leveraging the intrinsic connectivity of atom-based QPU modules, the architecture can be naturally extended to more complex network topologies for adaptive routing and multiuser entanglement distribution. Realizing such capabilities will require developing optimized routing strategies, resource-allocation algorithms, and scheduling policies for hybrid QPU-QM platforms.

In the current proposal, the loading of the entanglement is performed via additional entanglement swapping using a linear-optics Bell measurement. This is fundamentally probabilistic and imposes an overhead on the rate of the hybrid repeater~\cite{lutkenhaus1999bell}. The performance of the hybrid quantum repeater can potentially be further improved if a deterministic state transfer is used in the loading stage. One such possibility is to switch to a polarization-based double-click protocol~\cite{stephenson2020high, krutyanskiy2023entanglement} for remote-entanglement generation. Then existing cavity-assisted interaction can be used to map the polarization-encoded photons to atom-based qubits~\cite{duan2004scalable,reiserer2014quantum}. It would be interesting to investigate whether the double-click protocol with deterministic loading or the single-click protocol with probabilistic loading is optimal in realistic parameter regimes.

With continued advances in AFCQMs, integrated photonic interfaces, high-rate SPDC sources, and modular atom-based QPUs, we believe that the proposed hybrid repeater offers a realistic blueprint for building efficient first-generation quantum repeaters and for enabling scalable quantum networks of the future.

\emph{Note added.} We became aware of another work that also proposes a hybrid quantum repeater architecture~\cite{tissot2025hybrid}. Reference~\cite{tissot2025hybrid} and our proposal share similar motivation; however, the underlying architecture for interfacing QPU and AFCQM is different. In Ref.~\cite{tissot2025hybrid}, the ion-trap QPU output interferes directly with the SPDC photons, whereas in our protocol, the QPU performs entanglement swapping only after the heralding signal from the remote memory-memory entanglement is received. In our design, each elementary link results in QPU-QPU entanglement, enabling deterministic local quantum processing and simplifying the higher-level repeater scheduling. We also provide an explicit performance analysis of the hybrid architecture within a linear repeater-chain model. We believe that these distinctions, alongside our error mitigation study and repeater rate calculation model, reflect complementary perspectives on hybrid repeater design.

\begin{acknowledgments}
This work was supported by the Japan Science and Technology Agency (JST) Moonshot Research and Development Program (Grants No. JPMJMS2063 and No. JPMJMS226C). We thank T. Horikiri for insightful discussion.
\end{acknowledgments}

\section*{Data availability}
The scripts that can reproduce the numerical results of this paper are publicly available~\cite{sun2025hybrid_repeaters}.

\appendix

\section{State transformations in the hybrid repeater protocol}
\label{sec:appendix_state_change}

In this appendix, we derive the ideal state transformations during the hybrid repeater protocol in the absence of noise. We further assume perfect PNR detectors are used. The analytical forms for the density matrices where photon-loss noise is taken into account and/or the PNR assumption is dropped are presented in Appendix~\ref{sec:appendix_dm}. We use the following definition of Bell states for QPU-QPU entanglement:
\begin{align}
    |\Psi^+\rangle &= \sqrt{\frac{1}{2}}\left(|B_1 D_2\rangle + |D_1 B_2\rangle\right),
    \nonumber \\
    |\Psi^-\rangle &= \sqrt{\frac{1}{2}}\left(|B_1 D_2\rangle - |D_1 B_2\rangle\right).
\end{align}
In all equations, the number subscripts $\{1,2\}$ denote the left and right EGUs within an elementary link (Fig.~\ref{fig:hybrid_repeater}), and the operator $I$ denotes the identity operator acting on the state space indicated by its subscript.

We begin with the remote swapping step to generate QM-QM entanglement. Just prior to remote-entanglement swapping, the state after storage in the QMs [where idler modes are mapped to QM states according to Eq.~\eqref{eqn:joint_SPDC}] is given by
\begin{align}
    \ket{\psi_\text{SPDC,QM}} \propto \bigotimes^{N_\text{freq}}_{m = 1} \Big(
    &|0_{s_1,m} 0_{\text{QM1},m} 0_{s_2,m} 0_{\text{QM2},m}\rangle \nonumber \\
    {}+\lambda &|1_{s_1,m} 1_{\text{QM1},m} 0_{s_2,m} 0_{\text{QM2},m}\rangle \nonumber \\
    {}+\lambda &|0_{s_1,m} 0_{\text{QM1},m} 1_{s_2,m} 1_{\text{QM2},m}\rangle \nonumber \\
    {}+\lambda^2 &|2_{s_1,m} 2_{\text{QM1},m} 0_{s_2,m} 0_{\text{QM2},m}\rangle \nonumber \\
    {}+\lambda^2 &|0_{s_1,m} 0_{\text{QM1},m} 2_{s_2,m} 2_{\text{QM2},m}\rangle \nonumber \\
    {}+\lambda^2 &|1_{s_1,m} 1_{\text{QM1},m} 1_{s_2,m} 1_{\text{QM2},m}\rangle \Big),
\end{align}
where the subscripts QM1 and QM2 denote the QM states of the left and right EGUs, respectively.

\begin{figure}[tbp]
    \centering
    \includegraphics[width=0.8\linewidth]{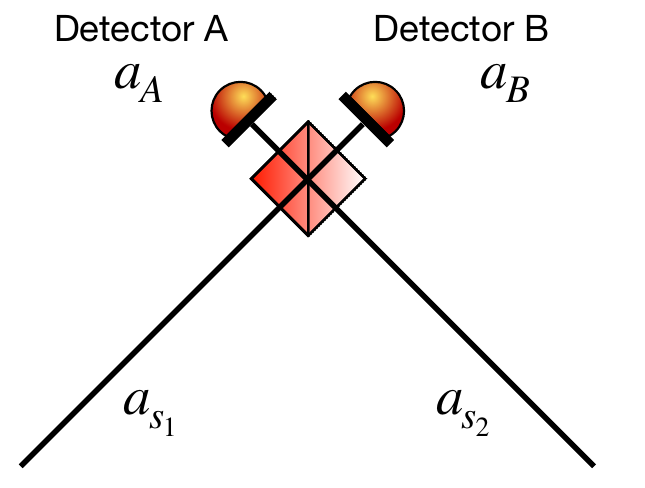}
    \caption{Illustration of the single-click protocol for remote-entanglement swapping. The operator $a_O$ corresponds to the annihilation operator for the specific mode indicated by the subscript $O$.}
    \label{fig:single_click}
\end{figure}

A successful heralding signal at the remote-entanglement swapper corresponds to a single detector click behind the beam splitter (Fig.~\ref{fig:single_click}). In the following analysis, we assume a successful heralding event is recorded for a particular frequency mode. We proceed with the heralded mode and omit the frequency mode index $m$ for brevity.

We define $a_O$ as the annihilation operator for optical mode $O$ and $|\text{vac}\rangle$ as the vacuum state ket vector. If detector $A$ clicks, the joint state is projected to
\begin{align}
    &(I_{\text{QM1},\text{QM2}} \otimes \langle \text{vac}| a_A) \ket{\psi_\text{SPDC,QM}}
    \nonumber \\
    &\quad\propto I_{\text{QM1},\text{QM2}} \otimes \left[\sqrt{\frac{1}{2}} \langle \text{vac}| (a_{s_1} + a_{s_2}) \right] \ket{\psi_\text{SPDC,QM}}
    \nonumber \\
    &\quad\propto \sqrt{\frac{1}{2}} I_{\text{QM1},\text{QM2}} \otimes (\langle 0_{s_1} 1_{s_2}| + \langle 1_{s_1} 0_{s_2}|) \ket{\psi_\text{SPDC,QM}}
    \nonumber \\
    &\quad\propto |0_\text{QM1}1_\text{QM2}\rangle + |1_\text{QM1} 0_\text{QM2}\rangle,
\end{align}
where in the first step we use the beam-splitter transformation to map the detector mode operators to the signal mode operators, using the convention $a_A =\sqrt{\frac{1}{2}} (a_{s_1} + a_{s_2})$ and $a_B =\sqrt{\frac{1}{2}} (a_{s_1} - a_{s_2})$.

Similarly, if detector $B$ clicks, then
\begin{align}
    &(I_{\text{QM1},\text{QM2}} \otimes \langle \text{vac}| a_B) \ket{\psi_\text{SPDC,QM}}
    \nonumber \\
    &\quad\propto I_{\text{QM1},\text{QM2}} \otimes \left[\sqrt{\frac{1}{2}} \langle \text{vac}| (a_{s_1} - a_{s_2}) \right] \ket{\psi_\text{SPDC,QM}}
    \nonumber \\
    &\quad\propto I_{\text{QM1},\text{QM2}} \otimes (\langle 0_{s_1} 1_{s_2}| - \langle 1_{s_1} 0_{s_2}|) \ket{\psi_\text{SPDC,QM}}
    \nonumber \\
    &\quad\propto -|0_\text{QM1}1_\text{QM2}\rangle + |1_\text{QM1} 0_\text{QM2}\rangle.
\end{align}
Therefore, both single-click events lead to a maximally entangled state up to a local phase. Without loss of generality, in the following discussion of single-click events, we consider only the positive-sign case, as the local phase can be corrected after the elementary link is generated.

Next the quantum memories retrieve the stored photons, and the loading of entanglement to the QPUs begins. The photonic state of the memory output modes in the Fock basis is
\begin{align}
    |\psi_\text{QM-out}\rangle &= \sqrt{\frac{1}{2}} \bigl(|0_\text{QM1-out}1_\text{QM2-out}\rangle \nonumber \\
    &\qquad{}+|1_\text{QM1-out}0_\text{QM2-out}\rangle\bigr),
\end{align}
where the subscripts QM1-out and QM2-out denote the output optical modes of the left and right QMs, respectively. In this stage, both QPUs generate atom-photon entangled states of the form given in Eq.~\eqref{eqn:ion_photon}, denoted by $|\psi_\text{QPU1,photon1}\rangle$ and $|\psi_\text{QPU2,photon2}\rangle$. We define the joint state just before loading as
\begin{align}
    |\psi_\text{load}\rangle = |\psi_\text{QPU1,photon1}\rangle \otimes |\psi_\text{QM-out}\rangle \otimes |\psi_\text{QPU2,photon2}\rangle.
\end{align}
We define the loading operator $P_\text{load} = P_\text{load,1} P_\text{load,2}$, where
\begin{align}
    P_\text{load,1} &= I_{\text{QPU1, QPU2, photon2, QM2-out}}
    \nonumber \\
    &\;\otimes \sqrt{\frac{1}{2}}
    \left(\langle 0_\text{photon1} 1_\text{QM1-out}| + \langle 1_\text{photon1} 0_\text{QM1-out} | \right),
    \nonumber \\
    P_\text{load,2} &= I_{\text{QPU1, QPU2, photon1, QM1-out}}
    \nonumber \\
    &\;\otimes \sqrt{\frac{1}{2}}
    \left(\langle 0_\text{photon2} 1_\text{QM2-out}| + \langle 1_\text{photon2} 0_\text{QM2-out} | \right).
\end{align}
The loading stage then transforms the joint state and establishes the QPU-QPU entanglement as follows:
\begin{align}
    P_\text{load} |\psi_\text{load}\rangle &\propto |\psi_\text{QPU-QPU}\rangle
    \nonumber \\
    &= \sqrt{\frac{1}{2}}\left(|B_1 D_2\rangle + |D_1 B_2\rangle \right).
\end{align}
Note that for the loading swappers, we only discussed click patterns leading to the $|\Psi^+\rangle$ state. Similar to the remote swapping case, other patterns exist leading to the $|\Psi^-\rangle$ state, which can always be corrected for.

\section{Analytical expressions for numerical density matrices in Sec.~\ref{sec:error_mitigation}}\label{sec:appendix_dm}
In this appendix, we provide analytical expressions for the density matrices in Eqs.~\eqref{eqn:EL_raw}, \eqref{eqn:EL_PNR}, \eqref{eqn:EL_EPL}, and \eqref{eqn:EL_RE} and the one corresponding to our proposed error-mitigation strategy ($\rho_\text{PNR+EPL}$). In the following expressions, we use $q$ for the atomic-emitter brightness, $\lambda$ for the SPDC brightness, $\varepsilon_r$ for the fiber-channel loss (remote loss), and $\varepsilon_l$ for the local loss. We use row-first and zero-based indexing for the density matrices as in Eq.~\eqref{eqn:EL_raw} and only show the lower-triangular part of each density matrix. All the terms are truncated at the order $\lambda^2$.

\begin{widetext}
\subsection{Case \texorpdfstring{$\rho_\text{raw}$}{rho raw}}
For this case we have

    \begin{align}
        \rho_\text{raw}[0,0]&=\frac{q^{2} \left[\lambda^{2} \left(19 \varepsilon_r \varepsilon_l^{2} + 20 \varepsilon_r \varepsilon_l + 13 \varepsilon_r + 9 \varepsilon_l^{2} + 12 \varepsilon_l + 7\right) + 8 \varepsilon_l + 8\right]}{N_\text{raw}},
        \nonumber \\
        \rho_\text{raw}[1,0] &= 0,
        \nonumber \\
        \rho_\text{raw}[1,1] &= \frac{q \left[\lambda^{2} \left(- 13 \varepsilon_r \varepsilon_l q + 13 \varepsilon_r \varepsilon_l - 7 \varepsilon_r q + 7 \varepsilon_r - 7 \varepsilon_l q + 7 \varepsilon_l - 5 q + 5\right) - 4 q + 4\right]}{N_\text{raw}},
        \nonumber \\
        \rho_\text{raw}[2,0] &= 0,
        \nonumber \\
        \rho_\text{raw}[2,1] &= 
        \frac{4 q \left[\lambda^{2} \left(- \varepsilon_r \varepsilon_l q + \varepsilon_r \varepsilon_l - \varepsilon_r q + \varepsilon_r - \varepsilon_l q + \varepsilon_l - q + 1\right) - q + 1\right]}{N_\text{raw}},
        \nonumber \\
        \rho_\text{raw}[2,2] &= 
        \rho_\text{raw}[1,1],
        \nonumber \\
        \rho_\text{raw}[3,0] &= 0,
        \nonumber \\
        \rho_\text{raw}[3,1] &= 0,
        \nonumber \\
        \rho_\text{raw}[3,2] &= 0,
        \nonumber \\
        \rho_\text{raw}[3,3] &= \frac{4 \lambda^{2} \left(\varepsilon_r q^{2} - 2 \varepsilon_r q + \varepsilon_r + q^{2} - 2 q + 1\right)}{N_\text{raw}},
        \nonumber \\
        N_\text{raw} &= \lambda^{2} \Bigl(19 \varepsilon_r \varepsilon_l^{2} q^{2} - 6 \varepsilon_r \varepsilon_l q^{2} + 26 \varepsilon_r \varepsilon_l q + 3 \varepsilon_r q^{2} + 6 \varepsilon_r q + 4 \varepsilon_r\nonumber \\
        &\quad{}+ 9 \varepsilon_l^{2} q^{2} - 2 \varepsilon_l q^{2} + 14 \varepsilon_l q + q^{2} + 2 q + 4\Bigr) + 8 \varepsilon_l q^{2} + 8 q.
    \end{align}

\subsection{Case \texorpdfstring{$\rho_\text{PNR}$}{rho PNR}}
For this case we have

    \begin{align}
    \rho_\text{PNR}[0,0] &= \frac{2 \varepsilon_l q^{2} \left(5 \lambda^{2} \varepsilon_r \varepsilon_l + 1\right)}{N_\text{PNR}},
    \nonumber \\
    \rho_\text{PNR}[1,0] &= 0, 
    \nonumber \\
    \rho_\text{PNR}[1,1] &=\frac{q \left[\lambda^{2} \left(- 8 \varepsilon_r \varepsilon_l q + 8 \varepsilon_r \varepsilon_l\right) - q + 1\right]}{2 N_\text{PNR}},
    \nonumber \\
    \rho_\text{PNR}[2,0] &= 0,
    \nonumber \\
    \rho_\text{PNR}[2,1] &= \frac{q \left[\lambda^{2} \left(- 4 \varepsilon_r \varepsilon_l q + 4 \varepsilon_r \varepsilon_l\right) - q + 1\right]}{2 N_\text{PNR}},
    \nonumber \\
    \rho_\text{PNR}[2,2] &=
    \rho_\text{PNR}[1,1],
    \nonumber \\
    \rho_\text{PNR}[3,0] &= 0,
    \nonumber \\
    \rho_\text{PNR}[3,1] &=0,
    \nonumber \\
    \rho_\text{PNR}[3,2] &=0,
    \nonumber \\
    \rho_\text{PNR}[3,3] &= \frac{\lambda^{2} \varepsilon_r \left(q^{2} - 2 q + 1\right)}{N_\text{PNR}},
    \nonumber \\
    N_\text{PNR} &= \lambda^{2} \left(10 \varepsilon_r \varepsilon_l^{2} q^{2} - 8 \varepsilon_r \varepsilon_l q^{2} + 8 \varepsilon_r \varepsilon_l q + \varepsilon_r q^{2} - 2 \varepsilon_r q + \varepsilon_r\right) + 2 \varepsilon_l q^{2} - q^{2} + q.
\end{align}

\subsection{Case \texorpdfstring{$\rho_\text{EPL}$}{rho EPL}}
For this case we write the final density matrix in terms of the matrix elements of $\rho_\text{raw}$:

\begin{align}
    \rho_\text{EPL}[0,0] &= \frac{\rho_\text{raw}[0,0] \rho_\text{raw}[3,3]}{N_\text{EPL}},
    \nonumber \\
    \rho_\text{EPL}[1,0] &= 0,
    \nonumber \\
    \rho_\text{EPL}[1,1] &= \frac{\rho_\text{raw}[1,1] \rho_\text{raw}[2,2]}{N_\text{EPL}},
    \nonumber \\
    \rho_\text{EPL}[2,0] &= 0,
    \nonumber \\
    \rho_\text{EPL}[2,1] &= \frac{\rho_\text{raw}[1,2] \rho_\text{raw}[2,1]}{N_\text{EPL}},
    \nonumber \\
    \rho_\text{EPL}[2,2] &= \frac{\rho_\text{raw}[1,1] \rho_\text{raw}[2,2]}{N_\text{EPL}},
    \nonumber \\
    \rho_\text{EPL}[3,0] &=
    0,
    \nonumber \\
    \rho_\text{EPL}[3,1] &= 0, 
    \nonumber \\
    \rho_\text{EPL}[3,2] &= 0,
    \nonumber \\
    \rho_\text{EPL}[3,3] &= \rho_\text{EPL}[0,0],
    \nonumber \\
    N_\text{EPL} &= 2 \rho_\text{raw}[0,0]
    \rho_\text{raw}[3,3] + 2 \rho_\text{raw}[1,1] \rho_\text{raw}[2,2].
\end{align}

\subsection{Case \texorpdfstring{$\rho_\text{RE}$}{rho RE}}
Similar to the $\rho_\text{EPL}$ case, we also write the final density matrix in terms of the matrix elements of $\rho_\text{raw}$:

    \begin{align}
        \rho_\text{RE}[0,0] &= \frac{8 \rho_\text{raw}[3,3] \varepsilon_l + 8 \rho_\text{raw}[3,3] + \lambda^{2} A}{N_\text{RE}},
        \nonumber \\
        \rho_\text{RE}[1,0] &= 0,
        \nonumber \\
        \rho_\text{RE}[1,1] &= \frac{4 \rho_\text{raw}[2,2] + \lambda^{2} \left(13 \rho_\text{raw}[2,2] \varepsilon_r \varepsilon_l + 7 \rho_\text{raw}[2,2] \varepsilon_r + 7 \rho_\text{raw}[2,2] \varepsilon_l + 5 \rho_\text{raw}[2,2]\right)}{N_\text{RE}},
        \nonumber \\
        \rho_\text{RE}[2,0] &= 0,
        \nonumber \\    
        \rho_\text{RE}[2,1] &= \frac{4 \rho_\text{raw}[1,2] + \lambda^{2} \left(4 \rho_\text{raw}[1,2] \varepsilon_r \varepsilon_l + 4 \rho_\text{raw}[1,2] \varepsilon_r + 4 \rho_\text{raw}[1,2] \varepsilon_l + 4 \rho_\text{raw}[1,2]\right)}{N_\text{RE}},
        \nonumber \\
        \rho_\text{RE}[2,2] &= \frac{4 \rho_\text{raw}[1,1] + \lambda^{2} \left(13 \rho_\text{raw}[1,1] \varepsilon_r \varepsilon_l + 7 \rho_\text{raw}[1,1] \varepsilon_r + 7 \rho_\text{raw}[1,1] \varepsilon_l + 5 \rho_\text{raw}[1,1]\right)}{N_\text{RE}},
        \nonumber \\
        \rho_\text{RE}[3,0] &= 0,
        \nonumber \\
        \rho_\text{RE}[3,1] &= 0,
        \nonumber \\
        \rho_\text{RE}[3,2] &= 0,
        \nonumber \\
        \rho_\text{RE}[3,3] &= \frac{\lambda^{2} \left(4 \rho_\text{raw}[0,0] \varepsilon_r + 4 \rho_\text{raw}[0,0]\right)}{N_\text{RE}},
        \nonumber \\
        N_\text{RE} &= 4 \rho_\text{raw}[1,1] + 4 \rho_\text{raw}[2,2] + 8 \rho_\text{raw}[3,3] \varepsilon_l + 8 \rho_\text{raw}[3,3] + \lambda^{2}B,
        \nonumber \\
        A &= 19 \rho_\text{raw}[3,3] \varepsilon_r \varepsilon_l^{2} + 20 \rho_\text{raw}[3,3] \varepsilon_r \varepsilon_l + 13 \rho_\text{raw}[3,3] \varepsilon_r + 9 \rho_\text{raw}[3,3] \varepsilon_l^{2} + 12 \rho_\text{raw}[3,3] \varepsilon_l + 7 \rho_\text{raw}[3,3],
        \nonumber \\
        B &= 4 \rho_\text{raw}[0,0] \varepsilon_r + 4 \rho_\text{raw}[0,0] + 13 \rho_\text{raw}[1,1] \varepsilon_r \varepsilon_l + 7 \rho_\text{raw}[1,1] \varepsilon_r + 7 \rho_\text{raw}[1,1] \varepsilon_l + 5 \rho_\text{raw}[1,1]  
        + 13 \rho_\text{raw}[2,2] \varepsilon_r \varepsilon_l 
        \nonumber \\
        &+ 7 \rho_\text{raw}[2,2] \varepsilon_r + 7 \rho_\text{raw}[2,2] \varepsilon_l + 5 \rho_\text{raw}[2,2] + 19 \rho_\text{raw}[3,3] \varepsilon_r \varepsilon_l^{2} + 20 \rho_\text{raw}[3,3] \varepsilon_r \varepsilon_l + 13 \rho_\text{raw}[3,3] \varepsilon_r 
        \nonumber \\
        &+ 9 \rho_\text{raw}[3,3] \varepsilon_l^{2} + 12 \rho_\text{raw}[3,3] \varepsilon_l + 7 \rho_\text{raw}[3,3].
    \end{align}

\end{widetext}

\subsection{Case \texorpdfstring{$\rho_\text{PNR+EPL}$}{rho PNR+EPL}}
For this case we have
\begin{align}
        \rho_\text{PNR+EPL}[0,0] &=4 \lambda^{2} \varepsilon_r \varepsilon_l + O(\lambda^4),
        \nonumber \\
        \rho_\text{PNR+EPL}[1,0] &= 0,
        \nonumber \\
        \rho_\text{PNR+EPL}[1,1] &= - 4 \lambda^{2} \varepsilon_r \varepsilon_l + \tfrac{1}{2} + O(\lambda^4),
        \nonumber \\
        \rho_\text{PNR+EPL}[2,0] &= 0,
        \nonumber \\
        \rho_\text{PNR+EPL}[2,1] &= - 8 \lambda^{2} \varepsilon_r \varepsilon_l + \tfrac{1}{2} + O(\lambda^4),
        \nonumber \\
        \rho_\text{PNR+EPL}[2,2] &= \rho_\text{PNR+EPL}[1,1],
        \nonumber \\
        \rho_\text{PNR+EPL}[3,0] &= 0,
        \nonumber \\
        \rho_\text{PNR+EPL}[3,1] &= 0,
        \nonumber \\
        \rho_\text{PNR+EPL}[3,2] &= 0,
        \nonumber \\
        \rho_\text{PNR+EPL}[3,3] &= \rho_\text{PNR+EPL}[0,0].
    \end{align}

\section{Rate-fidelity tradeoff for the hybrid repeater}
\label{sec:appendix_rate_fidelity}

\begin{figure}[tbp]
    \centering
    \includegraphics[width=\linewidth]{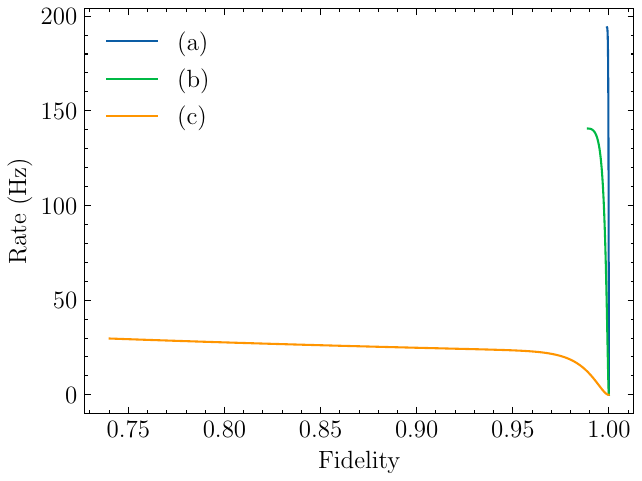}
    \caption{Rate-fidelity tradeoff for the hybrid repeater. Tradeoff frontiers are shown for various local-loss regimes: (a) $\eta_\text{QFC} = 0.99$ and $\eta_\text{QM} = 0.99$ (an idealized scenario representative of future devices), (b) $\eta_\text{QFC} = 0.9$ and $\eta_\text{QM} = 0.9$ (a near-term optimistic regime), and (c) $\eta_\text{QFC} = 0.5$ and $\eta_\text{QM} = 0.5$ (a conservative estimate based on current experimental parameters).}
    \label{fig:rate_fidelity}
\end{figure}

In this appendix, we present the rate-fidelity tradeoff for the hybrid quantum repeater. The results are calculated for a 10-km elementary link, with parameters set according to Table~\ref{tab:parameters}. We first perform a scan over the atom brightness ($0.01 \leq q \leq 0.99$) and the SPDC brightness ($0.01 \leq \lambda \leq 0.5$), recording the rate $r$ and fidelity $f$ at each point. The set of data $(f,r)$ collected in this way is first sorted in descending order of the fidelity. While running down this sorted list from the one with the highest fidelity, a new pair $(f',r')$ is taken as a point in the tradeoff frontier only if $r'$ is higher than all the previous rates. In this way, when the fidelity decreases along the tradeoff frontier, the rate only increases. As illustrated in Fig.~\ref{fig:rate_fidelity}, we show this tradeoff for three sets of local-loss parameters, identical to those in Fig.~\ref{fig:key_rates}. In both the ideal scenario [Fig.~\ref{fig:key_rates}(a)] and the near-term optimistic regime [Fig.~\ref{fig:key_rates}(b)], the fidelity on the tradeoff frontier remains high ($F\gtrsim 0.99$). On the other hand, under the conservative parameter regime [Fig.~\ref{fig:key_rates}(c)], the rate-fidelity tradeoff spans a much broader range of fidelity and achieves significantly lower maximum rates.

\section{Entanglement distribution rates}
\label{sec:appendix_rate}

\begin{figure*}[t]
    \centering
    \includegraphics[width=\linewidth]{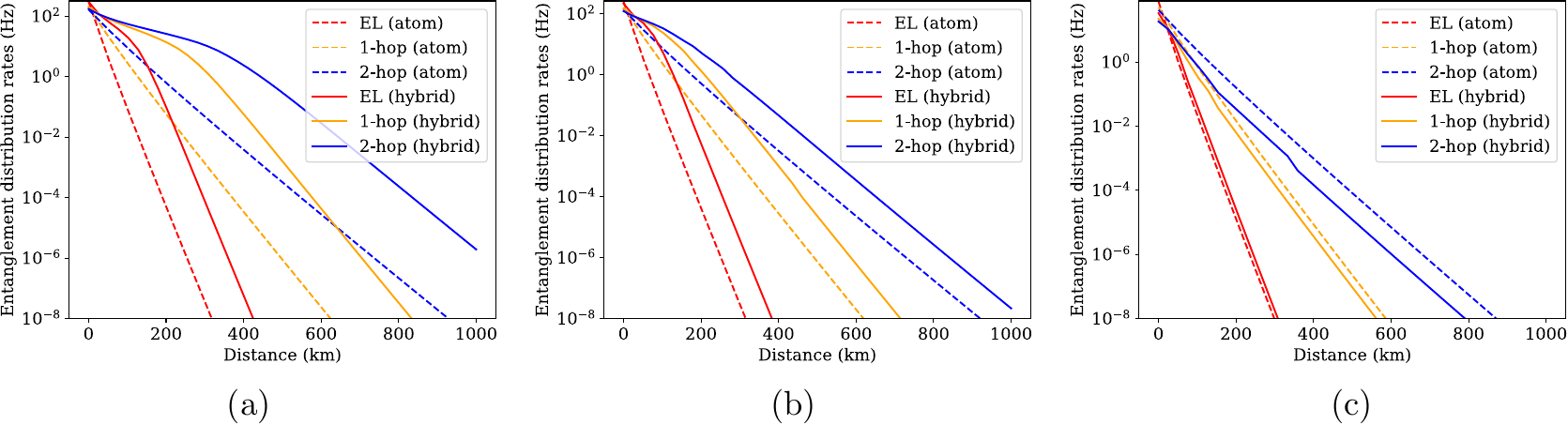}
    \caption{End-to-end entanglement distribution rates. The achievable entanglement distribution rate is plotted as a function of end-to-end distance with fidelity requirement $F\geq 0.95$. Dashed lines indicate the performance of atom-based repeater chains, while solid lines show the performance of the hybrid repeater. We evaluate three parameter regimes: (a) an idealized scenario representative of future devices ($\eta_\text{QFC} = 0.99$ and $\eta_\text{QM} = 0.99$), (b) a near-term optimistic regime ($\eta_\text{QFC} = 0.9$ and $\eta_\text{QM} = 0.9$), and (c) a regime based on conservative estimation of current experimental parameters ($\eta_\text{QFC} = 0.5$ and $\eta_\text{QM} = 0.5$).}
    \label{fig:entanglement_rate}
\end{figure*}

In this appendix, we present the raw entanglement-distribution rates obtained using the same parameters listed in Table~\ref{tab:parameters}. The reported rates correspond to the distribution of entangled pairs with fidelity at least $F = 0.95$; if this threshold is not met, the rate is recorded as zero. The results are shown in Fig.~\ref{fig:entanglement_rate}. Similar to the secret-key-rate analysis in Sec.~\ref{sec:results}, the proposed hybrid repeater outperforms the atom-based repeater in both the idealized [Fig.~\ref{fig:entanglement_rate}(a)] and the near-term optimistic [Fig.~\ref{fig:entanglement_rate}(b)] parameter regimes. In contrast, in the regime in Fig.~\ref{fig:entanglement_rate}(c), the more conservative parameter set, the atom-based architecture eventually surpasses the hybrid design as the repeater spacing increases. This behavior further highlights the importance of improving local hardware efficiencies and/or incorporating nested entanglement-distillation protocols to fully exploit the benefits of the hybrid architecture.

\section{Derivation of \texorpdfstring{$\langle t_\text{EL,D,all} \rangle$}{the mean tEL,D,all}}
\label{sec:appendix_t_EL_D_all}

Here we derive the expected time for all the elementary links between repeater stations to be generated such that the final deterministic entanglement swapping is ready to be performed.

We first consider an equivalent stochastic model for this problem. Assume we have an end-to-end connection divided into $N$ segments. Each segment has probability $p$ of becoming a connected link, and each trial round takes time $T$. In a given round, each segment attempts to connect independently. Let $X_i$ be the number of rounds until segment $i$ first succeeds. Then $X_i$ is geometric with probability $p$,
\begin{align}
    \Pr(X_i = k) = (1-p)^{k-1}p.
\end{align}
The tail probability of the geometric variable will be used,
\begin{align}
    \Pr(X_i>m) &= 1 - \sum_{n=1}^{m} \Pr (X_i = n)
    \nonumber \\
    &= (1-p)^m
\end{align}
for $m=0,1,2,\ldots$. This follows because not succeeding in the first $m$ rounds is equivalent to failing $m$ times. Equivalently,
\begin{align}
    \Pr(X_i \leq m) = 1-(1-p)^m.
\end{align}
Next we have an end-to-end connection only when all $N$ segments have succeeded at least once. The number of rounds until this happens is
\begin{align}
    R_N = \max\{X_1,\ldots,X_N\}.
\end{align}
For any integer $m \geq 0$,
\begin{align}
    \Pr(R_N \leq m) = \Pr(X_1 \leq m,\ldots,X_N \leq m).
\end{align}
Since each segment trial is performed independently, we have
\begin{align}
    \Pr(R_N \leq m) &= \prod_{i=1}^{N} \Pr(X_i \leq m) \nonumber \\
    &= [1-(1-p)^m]^N,
\end{align}
and the tail of $R_N$ is
\begin{align}
    \Pr(R_N > m) &= 1 - \Pr(R_N \leq m) \nonumber \\
    &= 1-[1-(1-p)^m]^N.
\end{align}
Next we invoke the tail-sum formula~\cite{grimmett2020probability} for a non-negative integer-valued random variable to calculate the expectation value of $R_N$.
\begin{align}
    \langle R_N \rangle &= \sum_{m=0}^\infty \Pr(R_N > m) \nonumber \\
    &= \sum_{m=0}^\infty [1 - \Pr(R_N \leq m)] \nonumber \\
    &= \sum_{m=0}^\infty \{1-[1-(1-p)^m]^N\}.
\end{align}
Since each round costs time $T$, the expected time is $T\langle R_N \rangle$. We can also turn the expression for $\langle R_N \rangle$ into a finite sum by using the binomial expansion. Letting $q = 1-p$, we have
\begin{align}
    (1-q^m)^N &= \sum_{j=0}^{N} \binom{N}{j} (-1)^j q^{jm}, \nonumber \\
    1-(1-q^m)^N &= \sum_{j=1}^{N} \binom{N}{j} (-1)^{j+1} q^{jm}.
\end{align}
Summing over $m \geq 0$ and swapping the sum leads to
\begin{align}
    \langle R_N \rangle &= \sum^{N}_{j=1}\binom{N}{j} (-1)^{j+1} \sum^{\infty}_{m=0} (q^j)^m \nonumber \\
    &= \sum_{j=1}^{N} \binom{N}{j} (-1)^{j+1} \frac{1}{1-q^j} \nonumber \\
    &= \sum^{N}_{j=1} (-1)^{j+1} \binom{N}{j} \frac{1}{1-(1-p)^j}.
\end{align}
The equation here becomes equivalent to Eq.~\eqref{eqn:t_EL_d_all} apart from the multiplication of $t_\text{EL,D}$ by setting $N = N_\text{QPU}-1$ and $p = p_\text{EL,D}$.


\begin{thebibliography}{65}

\bibitem{kimble2008quantum}
H. J. Kimble, The quantum internet, Nature (London) \textbf{453}, 1023 (2008).

\bibitem{wehner2018quantum}
S. Wehner, D. Elkouss, and R. Hanson, Quantum internet: A vision for the road ahead, Science \textbf{362}, eaam9288 (2018).

\bibitem{broadbent2009universal}
A. Broadbent, J. Fitzsimons, and E. Kashefi, Universal blind quantum computation, in \emph{Proceedings of the 2009 50th Annual IEEE Symposium on Foundations of Computer Science, Atlanta, 2009} (IEEE, Piscataway, NJ, 2009), pp. 517--526.

\bibitem{barz2012demonstration}
S. Barz, E. Kashefi, A. Broadbent, J. F. Fitzsimons, A. Zeilinger, and P. Walther, Demonstration of blind quantum computing, Science \textbf{335}, 303 (2012).

\bibitem{ekert1991quantum}
A. K. Ekert, Quantum cryptography based on Bell's theorem, Phys. Rev. Lett. \textbf{67}, 661 (1991).

\bibitem{ribordy2000long}
G. Ribordy, J. Brendel, J.-D. Gautier, N. Gisin, and H. Zbinden, Long-distance entanglement-based quantum key distribution, Phys. Rev. A \textbf{63}, 012309 (2000).

\bibitem{yin2017satellite}
J. Yin, Y. Cao, Y.-H. Li, J.-G. Ren, S.-K. Liao, L. Zhang, W.-Q. Cai, W.-Y. Liu, B. Li, H. Dai, \emph{et al.}, Satellite-to-ground entanglement-based quantum key distribution, Phys. Rev. Lett. \textbf{119}, 200501 (2017).

\bibitem{cirac1999distributed}
J. I. Cirac, A. K. Ekert, S. F. Huelga, and C. Macchiavello, Distributed quantum computation over noisy channels, Phys. Rev. A \textbf{59}, 4249 (1999).

\bibitem{van2010distributed}
R. Van Meter, T. D. Ladd, A. G. Fowler, and Y. Yamamoto, Distributed quantum computation architecture using semiconductor nanophotonics, Int. J. Quantum Inf. \textbf{08}, 295 (2010).

\bibitem{monroe2014large}
C. Monroe, R. Raussendorf, A. Ruthven, K. R. Brown, P. Maunz, L.-M. Duan, and J. Kim, Large-scale modular quantum-computer architecture with atomic memory and photonic interconnects, Phys. Rev. A \textbf{89}, 022317 (2014).

\bibitem{stephenson2020high}
L. J. Stephenson, D. P. Nadlinger, B. C. Nichol, S. An, P. Drmota, T. G. Ballance, K. Thirumalai, J. F. Goodwin, D. M. Lucas, and C. J. Ballance, High-rate, high-fidelity entanglement of qubits across an elementary quantum network, Phys. Rev. Lett. \textbf{124}, 110501 (2020).

\bibitem{krutyanskiy2023entanglement}
V. Krutyanskiy, M. Galli, V. Krcmarsky, S. Baier, D. A. Fioretto, Y. Pu, A. Mazloom, P. Sekatski, M. Canteri, M. Teller, \emph{et al.}, Entanglement of trapped-ion qubits separated by 230 meters, Phys. Rev. Lett. \textbf{130}, 050803 (2023).

\bibitem{covey2023quantum}
J. P. Covey, H. Weinfurter, and H. Bernien, Quantum networks with neutral atom processing nodes, npj Quantum Inf. \textbf{9}, 90 (2023).

\bibitem{bernien2013heralded}
H. Bernien, B. Hensen, W. Pfaff, G. Koolstra, M. S. Blok, L. Robledo, T. H. Taminiau, M. Markham, D. J. Twitchen, L. Childress, \emph{et al.}, Heralded entanglement between solid-state qubits separated by three metres, Nature (London) \textbf{497}, 86 (2013).

\bibitem{kalb2017entanglement}
N. Kalb, A. A. Reiserer, P. C. Humphreys, J. J. Bakermans, S. J. Kamerling, N. H. Nickerson, S. C. Benjamin, D. J. Twitchen, M. Markham, and R. Hanson, Entanglement distillation between solid-state quantum network nodes, Science \textbf{356}, 928 (2017).

\bibitem{yu2020entanglement}
Y. Yu, F. Ma, X.-Y. Luo, B. Jing, P.-F. Sun, R.-Z. Fang, C.-W. Yang, H. Liu, M.-Y. Zheng, X.-P. Xie, \emph{et al.}, Entanglement of two quantum memories via fibres over dozens of kilometres, Nature (London) \textbf{578}, 240 (2020).

\bibitem{liu2021heralded}
X. Liu, J. Hu, Z.-F. Li, X. Li, P.-Y. Li, P.-J. Liang, Z.-Q. Zhou, C.-F. Li, and G.-C. Guo, Heralded entanglement distribution between two absorptive quantum memories, Nature (London) \textbf{594}, 41 (2021).

\bibitem{briegel1998quantum}
H.-J. Briegel, W. D{\"u}r, J. I. Cirac, and P. Zoller, Quantum repeaters: The role of imperfect local operations in quantum communication, Phys. Rev. Lett. \textbf{81}, 5932 (1998).

\bibitem{munro2015inside}
W. J. Munro, K. Azuma, K. Tamaki, and K. Nemoto, Inside quantum repeaters, IEEE J. Sel. Top. Quantum Electron. \textbf{21}, 78 (2015).

\bibitem{azuma2023quantum}
K. Azuma, S. E. Economou, D. Elkouss, P. Hilaire, L. Jiang, H.-K. Lo, and I. Tzitrin, Quantum repeaters: From quantum networks to the quantum internet, Rev. Mod. Phys. \textbf{95}, 045006 (2023).

\bibitem{askarani2021entanglement}
M. F. Askarani, K. Chakraborty, and G. C. Do Amaral, Entanglement distribution in multi-platform buffered-router-assisted frequency-multiplexed automated repeater chains, New J. Phys. \textbf{23}, 063078 (2021).

\bibitem{gu2024hybrid}
F. Gu, S. G. Menon, D. Maier, A. Das, T. Chakraborty, W. Tittel, H. Bernien, and J. Borregaard, Hybrid quantum repeaters with ensemble-based quantum memories and single-spin photon transducers, npj Quantum Inf. \textbf{11}, 182 (2025).

\bibitem{cussenot2025uniting}
P. Cussenot, B. Grivet, L. Feldmann, S. Wengerowsky, B. P. Lanyon, T. E. Northup, H. de Riedmatten, A. S. S{\o}rensen, and N. Sangouard, Uniting quantum processing nodes of cavity-coupled ions with rare-earth quantum repeaters using single-photon pulse shaping based on atomic frequency comb, Phys. Rev. Lett. \textbf{135}, 240803 (2025).

\bibitem{afzelius2009multimode}
M. Afzelius, C. Simon, H. de Riedmatten, and N. Gisin, Multimode quantum memory based on atomic frequency combs, Phys. Rev. A \textbf{79}, 052329 (2009).

\bibitem{sinclair2014spectral}
N. Sinclair, E. Saglamyurek, H. Mallahzadeh, J. A. Slater, M. George, R. Ricken, M. P. Hedges, D. Oblak, C. Simon, W. Sohler, \emph{et al.}, Spectral multiplexing for scalable quantum photonics using an atomic frequency comb quantum memory and feed-forward control, Phys. Rev. Lett. \textbf{113}, 053603 (2014).

\bibitem{marcikic2002time}
I. Marcikic, H. de Riedmatten, W. Tittel, V. Scarani, H. Zbinden, and N. Gisin, Time-bin entangled qubits for quantum communication created by femtosecond pulses, Phys. Rev. A \textbf{66}, 062308 (2002).

\bibitem{takahashi2020strong}
H. Takahashi, E. Kassa, C. Christoforou, and M. Keller, Strong coupling of a single ion to an optical cavity, Phys. Rev. Lett. \textbf{124}, 013602 (2020).

\bibitem{li2024high}
Y. Li and J. D. Thompson, High-rate and high-fidelity modular interconnects between neutral atom quantum processors, PRX Quantum \textbf{5}, 020363 (2024).

\bibitem{sinclair2025fault}
J. Sinclair, J. Ramette, B. Grinkemeyer, D. Bluvstein, M. D. Lukin, and V. Vuleti{\'c}, Fault-tolerant optical interconnects for neutral-atom arrays, Phys. Rev. Res. \textbf{7}, 013313 (2025).

\bibitem{lago2021telecom}
D. Lago-Rivera, S. Grandi, J. V. Rakonjac, A. Seri, and H. de Riedmatten, Telecom-heralded entanglement between multimode solid-state quantum memories, Nature (London) \textbf{594}, 37 (2021).

\bibitem{hermans2023entangling}
S. L. N. Hermans, M. Pompili, L. Dos Santos Martins, A. R.-P. Montblanch, H. K. C. Beukers, S. Baier, J. Borregaard, and R. Hanson, Entangling remote qubits using the single-photon protocol: An in-depth theoretical and experimental study, New J. Phys. \textbf{25}, 013011 (2023).

\bibitem{bao2008generation}
X.-H. Bao, Y. Qian, J. Yang, H. Zhang, Z.-B. Chen, T. Yang, and J.-W. Pan, Generation of narrow-band polarization-entangled photon pairs for atomic quantum memories, Phys. Rev. Lett. \textbf{101}, 190501 (2008).

\bibitem{keller2004continuous}
M. Keller, B. Lange, K. Hayasaka, W. Lange, and H. Walther, Continuous generation of single photons with controlled waveform in an ion-trap cavity system, Nature (London) \textbf{431}, 1075 (2004).

\bibitem{ansari2018heralded}
V. Ansari, E. Roccia, M. Santandrea, M. Doostdar, C. Eigner, L. Padberg, I. Gianani, M. Sbroscia, J. M. Donohue, L. Mancino, M. Barbieri, and C. Silberhorn, Heralded generation of high-purity ultrashort single photons in programmable temporal shapes, Opt. Express \textbf{26}, 2764 (2018).

\bibitem{duan2001long}
L.-M. Duan, M. D. Lukin, J. I. Cirac, and P. Zoller, Long-distance quantum communication with atomic ensembles and linear optics, Nature (London) \textbf{414}, 413 (2001).

\bibitem{beukers2024remote}
H. K. C. Beukers, M. Pasini, H. Choi, D. Englund, R. Hanson, and J. Borregaard, Remote-entanglement protocols for stationary qubits with photonic interfaces, PRX Quantum \textbf{5}, 010202 (2024).

\bibitem{cabrillo1999creation}
C. Cabrillo, J. I. Cirac, P. Garcia-Fernandez, and P. Zoller, Creation of entangled states of distant atoms by interference, Phys. Rev. A \textbf{59}, 1025 (1999).

\bibitem{xiong2016active}
C. Xiong, X. Zhang, Z. Liu, M. J. Collins, A. Mahendra, L. G. Helt, M. J. Steel, D.-Y. Choi, C. J. Chae, P. H. W. Leong, \emph{et al.}, Active temporal multiplexing of indistinguishable heralded single photons, Nat. Commun. \textbf{7}, 10853 (2016).

\bibitem{schupp2021interface}
J. Schupp, V. Krcmarsky, V. Krutyanskiy, M. Meraner, T. E. Northup, and B. P. Lanyon, Interface between trapped-ion qubits and traveling photons with close-to-optimal efficiency, PRX Quantum \textbf{2}, 020331 (2021).

\bibitem{krovi2016practical}
H. Krovi, S. Guha, Z. Dutton, J. A. Slater, C. Simon, and W. Tittel, Practical quantum repeaters with parametric downconversion sources, Appl. Phys. B \textbf{122}, 52 (2016).

\bibitem{grimau2017heralded}
M. Grimau Puigibert, G. H. Aguilar, Q. Zhou, F. Marsili, M. D. Shaw, V. B. Verma, S. W. Nam, D. Oblak, and W. Tittel, Heralded single photons based on spectral multiplexing and feed-forward control, Phys. Rev. Lett. \textbf{119}, 083601 (2017).

\bibitem{rielander2017frequency}
D. Riel{\"a}nder, A. Lenhard, O. J. Far{\`i}as, A. M{\'a}ttar, D. Cavalcanti, M. Mazzera, A. Ac{\'i}n, and H. de Riedmatten, Frequency-bin entanglement of ultra-narrow band non-degenerate photon pairs, Quantum Sci. Technol. \textbf{3}, 014007 (2018).

\bibitem{guha2015rate}
S. Guha, H. Krovi, C. A. Fuchs, Z. Dutton, J. A. Slater, C. Simon, and W. Tittel, Rate-loss analysis of an efficient quantum repeater architecture, Phys. Rev. A \textbf{92}, 022357 (2015).

\bibitem{bonarota2011highly}
M. Bonarota, J.-L. Le Gou{\"e}t, and T. Chaneliere, Highly multimode storage in a crystal, New J. Phys. \textbf{13}, 013013 (2011).

\bibitem{wei2024quantum}
S.-H. Wei, B. Jing, X.-Y. Zhang, J.-Y. Liao, H. Li, L.-X. You, Z. Wang, Y. Wang, G.-W. Deng, H.-Z. Song, \emph{et al.}, Quantum storage of 1650 modes of single photons at telecom wavelength, npj Quantum Inf. \textbf{10}, 19 (2024).

\bibitem{england2015storage}
D. G. England, K. A. G. Fisher, J.-P. W. MacLean, P. J. Bustard, R. Lausten, K. J. Resch, and B. J. Sussman, Storage and retrieval of THz-bandwidth single photons using a room-temperature diamond quantum memory, Phys. Rev. Lett. \textbf{114}, 053602 (2015).

\bibitem{merkouche2022heralding}
S. Merkouche, V. Thiel, A. O. C. Davis, and B. J. Smith, Heralding multiple photonic pulsed Bell pairs via frequency-resolved entanglement swapping, Phys. Rev. Lett. \textbf{128}, 063602 (2022).

\bibitem{hong1987measurement}
C.-K. Hong, Z.-Y. Ou, and L. Mandel, Measurement of subpicosecond time intervals between two photons by interference, Phys. Rev. Lett. \textbf{59}, 2044 (1987).

\bibitem{cahall2017multi}
C. Cahall, K. L. Nicolich, N. T. Islam, G. P. Lafyatis, A. J. Miller, D. J. Gauthier, and J. Kim, Multi-photon detection using a conventional superconducting nanowire single-photon detector, Optica \textbf{4}, 1534 (2017).

\bibitem{kawasaki2022generation}
A. Kawasaki, K. Takase, T. Nomura, S. Miki, H. Terai, M. Yabuno, F. China, W. Asavanant, M. Endo, J.-i. Yoshikawa, \emph{et al.}, Generation of highly pure single-photon state at telecommunication wavelength, Opt. Express \textbf{30}, 24831 (2022).

\bibitem{endo2021quantum}
M. Endo, T. Sonoyama, M. Matsuyama, F. Okamoto, S. Miki, M. Yabuno, F. China, H. Terai, and A. Furusawa, Quantum detector tomography of a superconducting nanostrip photon-number-resolving detector, Opt. Express \textbf{29}, 11728 (2021).

\bibitem{nickerson2014freely}
N. H. Nickerson, J. F. Fitzsimons, and S. C. Benjamin, Freely scalable quantum technologies using cells of 5-to-50 qubits with very lossy and noisy photonic links, Phys. Rev. X \textbf{4}, 041041 (2014).

\bibitem{barrett2005efficient}
S. D. Barrett and P. Kok, Efficient high-fidelity quantum computation using matter qubits and linear optics, Phys. Rev. A \textbf{71}, 060310(R) (2005).

\bibitem{bennett1992quantum}
C. H. Bennett, G. Brassard, and N. D. Mermin, Quantum cryptography without Bell's theorem, Phys. Rev. Lett. \textbf{68}, 557 (1992).

\bibitem{scarani2009security}
V. Scarani, H. Bechmann-Pasquinucci, N. J. Cerf, M. Du{\v{s}}ek, N. L{\"u}tkenhaus, and M. Peev, The security of practical quantum key distribution, Rev. Mod. Phys. \textbf{81}, 1301 (2009).

\bibitem{monz2009realization}
T. Monz, K. Kim, A. S. Villar, P. Schindler, M. Chwalla, M. Riebe, C. F. Roos, H. H{\"a}ffner, W. H{\"a}nsel, M. Hennrich, and R. Blatt, Realization of universal ion-trap quantum computation with decoherence-free qubits, Phys. Rev. Lett. \textbf{103}, 200503 (2009).

\bibitem{bruzewicz2019trapped}
C. D. Bruzewicz, J. Chiaverini, R. McConnell, and J. M. Sage, Trapped-ion quantum computing: Progress and challenges, Appl. Phys. Rev. \textbf{6}, 021314 (2019).

\bibitem{myerson2008high}
A. H. Myerson, D. J. Szwer, S. C. Webster, D. T. C. Allcock, M. J. Curtis, G. Imreh, J. A. Sherman, D. N. Stacey, A. M. Steane, and D. M. Lucas, High-fidelity readout of trapped-ion qubits, Phys. Rev. Lett. \textbf{100}, 200502 (2008).

\bibitem{dur1999quantum}
W. D{\"u}r, H.-J. Briegel, J. I. Cirac, and P. Zoller, Quantum repeaters based on entanglement purification, Phys. Rev. A \textbf{59}, 169 (1999).

\bibitem{lutkenhaus1999bell}
N. L{\"u}tkenhaus, J. Calsamiglia, and K.-A. Suominen, Bell measurements for teleportation, Phys. Rev. A \textbf{59}, 3295 (1999).

\bibitem{duan2004scalable}
L.-M. Duan and H. J. Kimble, Scalable photonic quantum computation through cavity-assisted interactions, Phys. Rev. Lett. \textbf{92}, 127902 (2004).

\bibitem{reiserer2014quantum}
A. Reiserer, N. Kalb, G. Rempe, and S. Ritter, A quantum gate between a flying optical photon and a single trapped atom, Nature (London) \textbf{508}, 237 (2014).

\bibitem{tissot2025hybrid}
B. Tissot, S. Maiti, E. R. Hellebek, and A. S. S{\o}rensen, Hybrid single-ion atomic-ensemble node for high-rate remote entanglement generation, Phys. Rev. A \textbf{114}, L010602 (2026).

\bibitem{sun2025hybrid_repeaters}
S. Sun, hybrid repeaters, GitHub, San Francisco, 2025, \url{https://github.com/zinc-sun/hybrid_repeater}.

\bibitem{grimmett2020probability}
G. Grimmett and D. Stirzaker, \emph{Probability and Random Processes} (Oxford University Press, Oxford, 2020).

\end{thebibliography}
\end{document}